\documentclass[10pt,english]{article}
\usepackage{mathptmx}
\usepackage[T1]{fontenc}
\usepackage[latin9]{inputenc}
\usepackage{geometry}
\usepackage{color}
\usepackage{babel}
\usepackage{float}
\usepackage{amsmath}
\usepackage{amsthm}
\usepackage{graphicx}
\usepackage{esint}
\usepackage[authoryear]{natbib}
\usepackage[unicode=true,pdfusetitle,
 bookmarks=true,bookmarksnumbered=true,bookmarksopen=true,bookmarksopenlevel=13,
 breaklinks=false,pdfborder={0 0 0},pdfborderstyle={},backref=false,colorlinks=true]
 {hyperref}

\usepackage{lineno}
\usepackage{babel}
\usepackage{amsthm}
\usepackage{mathptmx}
\usepackage{stackrel}
\usepackage{graphicx}
\usepackage{esint}
\usepackage{babel}
\usepackage{units}
\usepackage{mathrsfs}
\usepackage{amsmath}
\usepackage{graphicx}
\usepackage{setspace}
\usepackage[authoryear]{natbib}
\usepackage{soul}
\usepackage{geometry}
\makeatletter
\newcommand{\lyxaddress}[1]{
	\par {\raggedright #1
	\vspace{1.4em}
	\noindent\par}
}

\@ifundefined{date}{}{\date{}}
\usepackage{color}
\usepackage[usenames,dvipsnames,svgnames,table]{xcolor}
\usepackage{multicol}
\definecolor{burgundy}{rgb}{0.5, 0.0, 0.13}
\definecolor{airforceblue}{rgb}{0.36, 0.54, 0.66}
\hypersetup{urlcolor=airforceblue}
\hypersetup{citecolor=burgundy}

\setcitestyle{notesep={, },super,open={},close={},comma,aysep={,},yysep={,}}  % this doesn't
 \makeatletter
   \renewcommand\@biblabel[1]{#1.}
 \makeatother

\makeatother

\begin{document}
\title{The oscillatory motion of Jupiter's polar cyclones results from vorticity dynamics}
\author{\href{https://orcid.org/0000-0002-3645-0383}{Nimrod Gavriel}$^{1\star}$
and \href{https://orcid.org/0000-0003-4089-0020}{Yohai Kaspi}$^{1}$}
\maketitle

\lyxaddress{\begin{center}
\textit{$^{1}$Department of Earth and Planetary Sciences, Weizmann
Institute of Science, Rehovot, Israel}\\
\textit{$^{\star}$\href{mailto:nimrod.gavriel@weizmann.ac.il}{nimrod.gavriel@weizmann.ac.il}}
\par\end{center}}

\lyxaddress{\begin{center}
Preprint June 8, 2022\textit{}\\
\textit{Geophys. Res. Lett.}. 49, 15, (2022). DOI:\href{https://doi.
org/10.1029/2022GL098708}{10.1029/2022GL098708}
\\
(Received Mars 22, 2022; Revised June 4, 2022; Accepted June
10, 2022) 
\par\end{center}}
\begin{abstract}
The polar cyclone at Jupiter's south pole and the 5 cyclones surrounding it oscillate in position and interact. These cyclones, observed since 2016 by NASA's Juno mission, present a unique opportunity to study vortex dynamics and interactions on long time scales. The cyclones' position data, acquired by Juno's JIRAM instrument, is analyzed, showing dominant oscillations with $\sim$12-month periods and amplitudes of $\sim400$ km. Here, the mechanism driving these oscillations is revealed by considering vorticity-gradient forces generated by mutual interactions between the cyclones and the latitudinal variation in planetary vorticity. Data-driven estimation of these forces exhibits a high correlation with the measured acceleration of the cyclones. To further test this mechanism, a model is constructed, simulating how cyclones subject to these forces exhibit similar oscillatory motion. 
\end{abstract}
%TC:ignore

%TC:endignore

\twocolumn 

\subsection*{Plain Language Summary}
The poles of Jupiter, observed by NASA's Juno spacecraft since 2016, have a unique symmetric arrangement of storms, where a polar cyclone situated at each pole is surrounded by eight cyclones in the north pole and five cyclones at the south pole. These cyclones, traced for over five years, move with a circular periodic pattern with a ~1-year cycle. Here, we explain these periodic patterns by considering the sum of the mutual rejection forces and the polar attraction force resulting from the conservation of momentum inside each cyclone. Using the cyclones' position data, a calculation of these forces agrees with their measured acceleration, supporting the suggested mechanism. In addition, a simplified model is constructed, which simulates how the south polar cyclones would move under these forces and predict the same type of circular motion. 
\subsection*{Key Points:}
\begin{itemize}
\item {The cyclones at Jupiter's south pole oscillate around their equilibrium positions with a $\sim$1 year period and a $\sim$400 km amplitude}
\item {A poleward beta-drift force and mutual rejection forces between the cyclones are established as the drivers of the observed oscillations}
\item {The suggested mechanism is supported both by calculations of accelerations and forces based on observations and by an ideal model
simulation}
\end{itemize}

\section{Introduction}
\begin{figure}
\begin{centering}
\includegraphics[width=1\columnwidth]{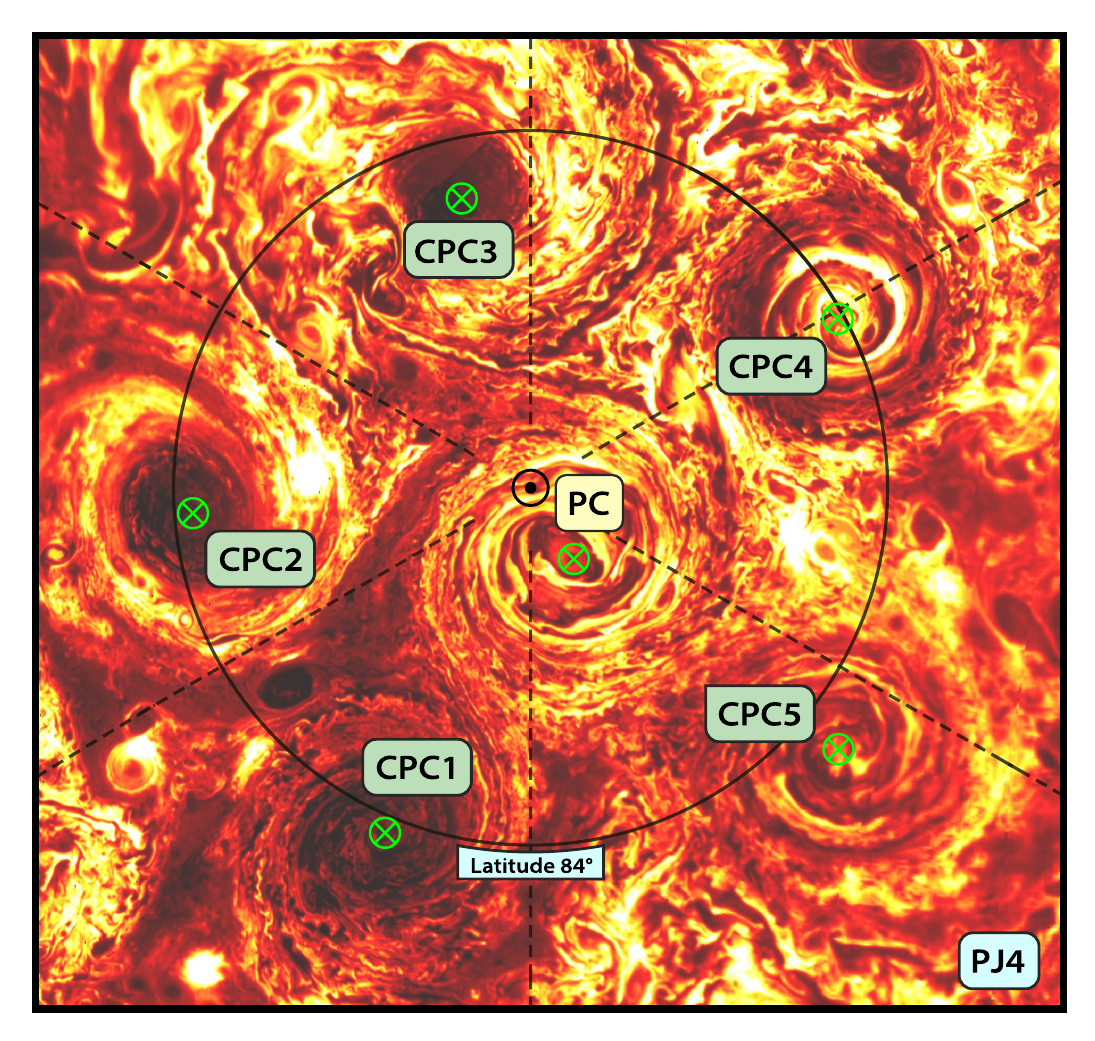}
\par\end{centering}
\caption{Infrared radiation photograph of Jupiter's south pole taken by the Jovian Auroral Mapper (JIRAM) instrument on Juno.
Image credit: NASA/JPL-Caltech/SwRI/ASI/INAF/JIRAM. The black point
represents the pole. The dashed lines represent intervals of $60^{\circ}$
longitude (longitude $0^{\circ}$ in System III is the line going
\textquotedblleft upward\textquotedblright{} from the pole). The green
'X's are the estimated location at the centers of the south polar
cyclones during PJ4. \label{fig: Photo}}
\end{figure}

\begin{figure*}[ht!]
\begin{centering}
\includegraphics[width=0.9\textwidth]{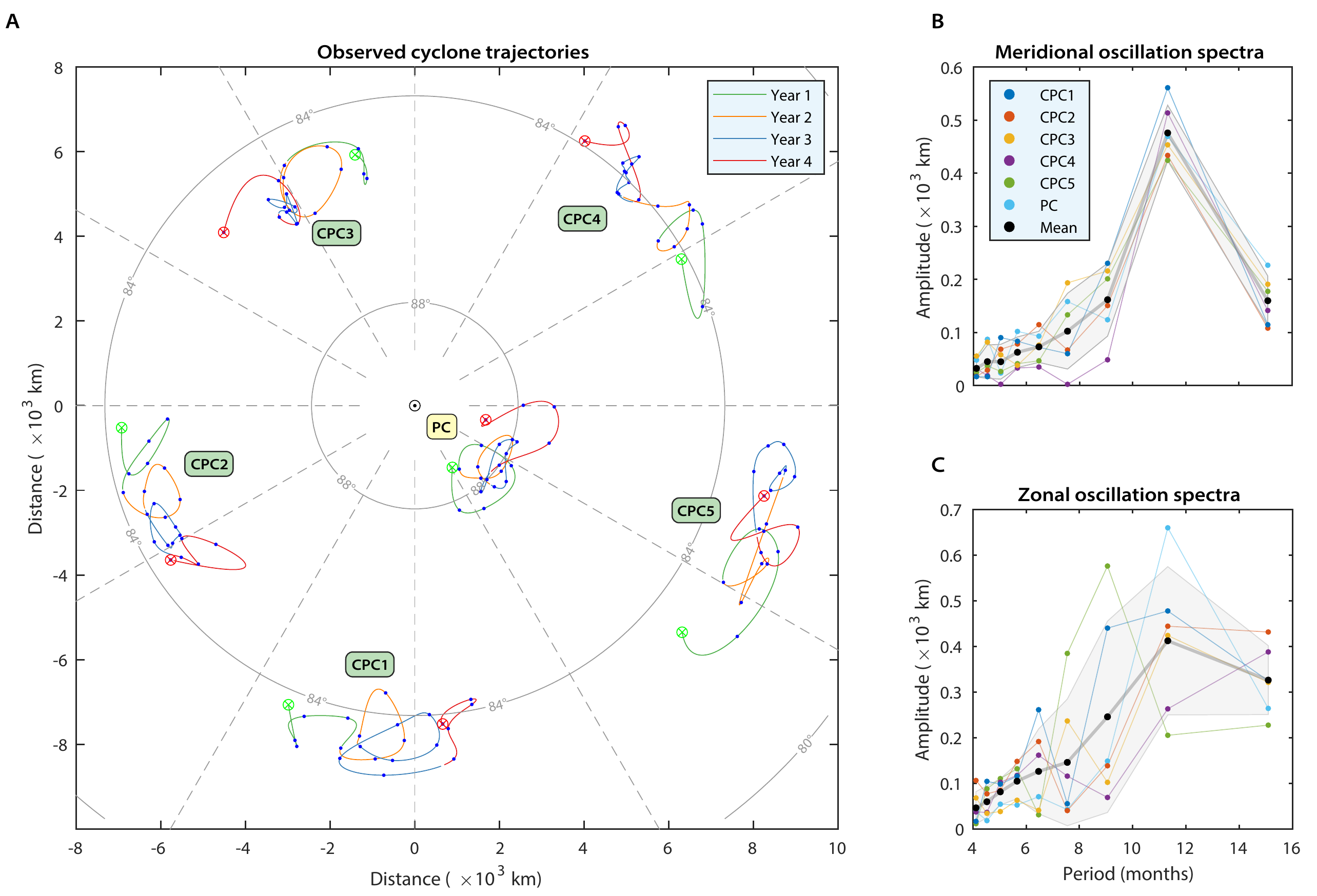}
\par\end{centering}
\caption{(a) The observed trajectories of the cyclones
at the south poles. The green (red) 'X's represent the locations of
the cyclones at PJ4 (30), which are the initial (final) locations in
the tracked period. For context, see JIRAM's image from PJ4 (Fig. \ref{fig: Photo}).
Dots represent observed coordinates \cite{mura2021oscillations}. The lines are cubic "spline" interpolations between the dots, and are therefore speculative. The lines change their
color in the passing of every year since PJ4 (2nd of Feb. 2017).
As the data contain less than 4 consecutive years, the red curves
only represent 9 months. The black dot represents the pole, gray circles
are latitude {[}$^{\circ}${]}, and dashed gray lines represent intervals
of $30^{\circ}$ longitude (longitude $0^{\circ}$ in System III is
the line going \textquotedblleft upward\textquotedblright{} from the
pole). See Movie~S1 for an animation of these trajectories. See Fig.\:\ref{fig: 1 time_series} for time series of the motion's longitude and latitude. (c,d) The frequency spectra of the observed meridional and
zonal (interpolated) motion of the cyclones, respectively. Straight
colored lines are linear interpolations between the calculated values
(colored dots) and thus do not represent real spectra. Black dots
represent the mean spectra between the 6 cyclones, and the gray shade
represents a standard deviation (calculated between the 6 cyclones)
around the mean. These spectra is presented in terms of oscillation energy in Fig.\:\ref{fig: 1 power spectra}. Notice that the time units presented in this study
are Earth's years and months and therefore do not hold physical significance
such as orbital periods.\label{fig:a.Data}}
\end{figure*}
The poles of Jupiter were observed in detail for the first time by
the Juno spacecraft in 2016 \cite{bolton2017jupiter}. In contrast
with the polar regions of Saturn, which are inhabited by a single
polar cyclone (PC) \cite{sanchez2006strong,baines2009saturn}, Jupiter's
PCs are surrounded by a ring of stable circumpolar cyclones (CPCs)
\cite{orton2017first,adriani2018}. There are 8 CPCs at the north
pole and five at the south (Fig. \ref{fig: Photo}), each with a diameter
of roughly $5,000$~km and velocities reaching $100$ ms$^{-1}$
\cite{adriani2020two,grassi2018first}. Such cyclones can be generated
by moist convection \cite{oneill2015polar,oneill2016}, where 2D
inverse energy cascade in the turbulent polar regions brings the kinetic
energy from the convective scale up to the horizontal scale of the
cyclones \cite{moriconi2020turbulence,siegelman2022moist}. These
regions are bounded by prograde jets at around latitudes $65^{\circ}$N\textbackslash S
\cite{rogers2022flow,rogers2017jupiter}, which may act as a separating
barrier. In contrast with the Great Red Spot, which is centered around latitude $20^{\circ}$S and
has a shallow depth (less than $500$~km \citep{parisi2021depth})
relative to the deep surrounding jets ($\sim3000$~km\citep{kaspi2018jupiter}),
the polar cyclones, subject to the Taylor-Proudman theorem \cite{vallis2017atmospheric}
with a vertical axis parallel to the planetary rotation axis and a smaller
Rossby number, potentially extend deeper, suggesting a 2D dynamical
regime.

The beta-drift is a force that results from a dipole of vorticity
(usually termed ``beta-gyres''\citep{sutyrin1994intense}) that
is induced by an interaction between the tangential velocity of a
cyclone and a gradient of potential vorticity (PV) that is present
in the background of the cyclone \cite{rossby1948displacements,chan2005physics,gavriel2021number}.
This force creates a poleward drift on cyclones and an equatorward
drift on anticyclones when only the planetary vorticity gradient ($\beta$) is
considered \cite{franklin1996tropical,chan2005physics,merlis2019aquaplanet}.
Beta-drift is a known contributor to the poleward motion of Earth's
tropical cyclones \cite{zhao2009observational}, and was shown in
shallow-water (SW) models to result in cyclones merging into a PC
in settings characterizing gas-giants such as Jupiter and Saturn \cite{scott2011polar,oneill2016,brueshaber2019dynamical,brueshaber2021effects}. Also, vortex crystals\citep{fine1995relaxation,schecter1999vortex} similar to Jupiter's were shown to form around the pole from turbulent initial conditions under the influence of $\beta$ in quasi-geostrophic (QG) model simulations\citep{siegelman2022polar}. However, other theories for poleward drift of cyclones exist \cite{afanasyev2018cyclonic,afanasyev2020poleward}.
 In addition to $\beta$, any PV
gradient in the background of a cyclone (e.g., by a jet or a nearby
cyclone) can similarly influence its motion \cite{chan1995interaction,shin2006critical,Riviere2012}. For example, the PV gradient due to the zonal mean wind shear and $\beta$ was shown, using full 3D models, to drive the meridional motion of large vortices on Uranus \cite{lebeau2020numerical} and Neptune \cite{lebeau1998epic}.
It was recently found that latitude $\sim84^{\circ}$, where the CPCs
are observed at both poles of Jupiter, is the latitude where the beta-drift
is balanced by an equivalent vorticity-gradient force, in which
the source of the PV gradient field is the PC instead of the planetary
$\beta$ \cite{gavriel2021number}. Such a rejection force between
cyclones is dependent on them having an anticyclonic ring around them
\cite{Li2020}, often referred to a 'vortex shielding', which is consistent with observations \cite{grassi2018first,gavriel2021number,ingersoll2021polygonal}
and SW simulations of the polar cyclones \cite{Li2020}. This balance
between the forces proved impossible on Saturn, as its PCs are too
large, such that the annulus where they could reject CPCs is too far
from the poles, where $\beta$ dominates, thus inhibiting Saturnian
CPCs \cite{gavriel2021number}.

The locations of the CPCs have been monitored since Juno's arrival
at Jupiter \cite{mura2021oscillations,adriani2020two,tabataba2020long}.
While the average locations of the cyclones match
the calculations of Gavriel and Kaspi (2021)\cite{gavriel2021number} (hereafter
GK21), the individual cyclones were found to oscillate around these stable positions,
where perturbations seem to pass on between neighboring cyclones \cite{mura2021oscillations}.
In addition, an average westward drift of approximately $7.5^{\circ}$
($3^{\circ}$) per year was measured on the cyclones at the south
(north) pole \cite{adriani2020two,mura2021oscillations}. Here, we
extend the work laid out by GK21 and show that in addition to determining
the equilibrium conditions of the CPCs, vorticity-gradient forces
also control their observed oscillatory motion.

\section{Results}

\subsection{Characterizing the Motion of the Cyclones in the Data }
\begin{figure*}[th!]
\begin{centering}
\includegraphics[width=0.7\textwidth]{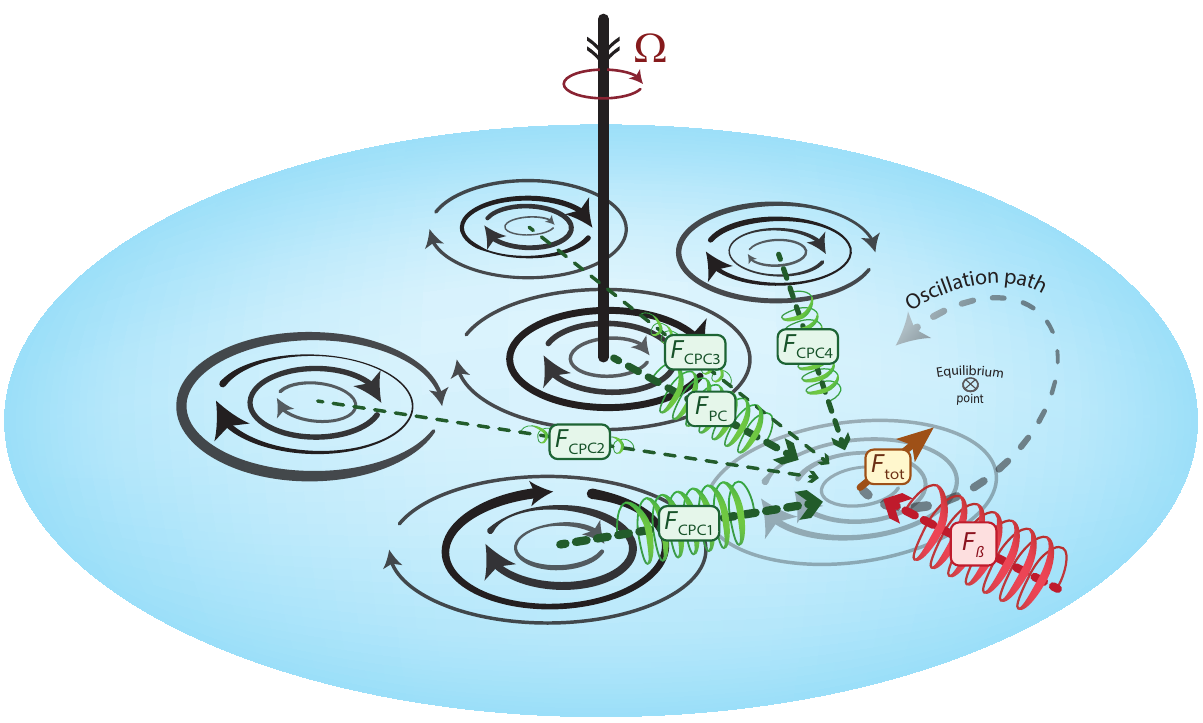}
\par\end{centering}
\caption{An illustration of the dynamical process driving the observed
motion of the south polar cyclones. The thick black line in the center
represents the planetary rotation axis and points to the pole. The
green arrows and springs represent the rejection forces acting on
a CPC due to the vorticity-gradient forces by the individual cyclones. The direction of these forces might rotate by a small factor, as described in section 2.2.
The magnitude of the forces ($F_{{\rm CPC}i}$ by CPC number $i$,
and $F_{{\rm PC}}$ by the PC), expressed qualitatively by the size
of the arrows, decreases with the distance between the cyclones.
Therefore, the far CPC2 and CPC3 exert minor rejection on the forced
CPC. The red $F_{\beta}$ represents the polar attraction due to the
beta-drift. The blue shade represents the potential magnitude of $F_{\beta}$
as a function of latitude, which vanishes (white
shade) at the pole and gets larger (bluer shades) away from it. The
brown arrow ($F_{{\rm tot}}$) represents the total net force acting
on the cyclone. $F_{{\rm tot}}$ points towards the equilibrium point,
the location where the net force vanishes. The gray dashed arrow illustrates
an oscillatory path around the equilibrium point. Due to the motion's
inertia, the net force is perpendicular to the path. Note that the portrayed forces are more complex than a linear spring (e.g., they only "push" away from the source and never "pull" toward it), and thus the representation of them as springs is made only for illustrative purposes.  \label{fig:Theory}}
\end{figure*}

To analyze the motion of the PC and the CPCs, we use the data gathered
by Juno's JIRAM instrument \cite{mura2021oscillations}. As the measurements
of the cyclones at the north pole were relatively infrequent, we make
the analyses only for the south pole. Each orbit of Juno is termed
a perijove (PJ), taking \textasciitilde 53 days per orbit during
the first 30 PJs analyzed in this study. An interpolation of the path
that the cyclones went through between PJs is shown in Fig.~\ref{fig:a.Data}a,
where the trajectories of the cyclones are divided by color to the
3 years and 9 months of observations between PJs 4 and 30. Considering
for example CPC2, it can be seen that the motion of the cyclone can
be inferred as a circular $~\sim$1-year oscillation with an amplitude
of $\sim3^{\circ}$ longitude (or $\sim0.3^{\circ}$ latitude), orbiting
an equilibrium point that migrates westward at $\sim7^{\circ}$ longitude
per year.

We suggest here that this motion results from a dynamical system of
6 bodies (cyclones) with 15 spring-like connections (i.e., rejection
forces as described in GK21) between them, in addition to the beta-drift
polar attraction acting independently on each cyclone. This picture
is illustrated for one cyclone in Fig.~\ref{fig:Theory}, where the
subject cyclone is rejected by the 5 other cyclones, where each rejection
force decreases in magnitude with the distance between the cyclones,
and is pushed toward the pole by the beta-drift force ($F_{\beta}$).
The magnitude of $F_{\beta}$ increases with distance from the pole
\cite{gavriel2021number}. The net force ($F_{{\rm tot}}$), resulting
from the summation of the 6 forces, deflects the trajectory of the
cyclone and results in a circular path concentric to the position
of equilibrium. This scheme gets more complicated when taking into
account that the 5 rejecting cyclones in Fig.~\ref{fig:Theory} move
and oscillate as well, resulting in a more chaotic pattern, which
includes various modes between the 6 cyclones. These interferences
can be seen in the more complex paths of the 6 cyclones (Fig.~\ref{fig:a.Data}a).
Note that the cumulative trajectory of the cyclones (Fig.~\ref{fig:a.Data}a)
is not entirely concentric to the pole, but is displaced. This can
happen, in the context of the suggested mechanism, if CPCs 2 and 3
``weigh'' more than the other cyclones in terms of their rejection
forces and their attraction to the pole, thus they get closer poleward
while pushing the other cyclones father away.

To further dissect the observed motion, it is informative to look
on the spectra of the observed motion in the meridional and zonal
directions (Fig.~\ref{fig:a.Data}b-c). All of the cyclones have
a dominant oscillation mode of a $\sim1$-year period and $\sim$400~km
amplitude. In the zonal direction (Fig.~\ref{fig:a.Data}c), the
spectrum is more dispersed, perhaps due to the absence of the regulating
effect of the beta-drift in the zonal direction, leaving only the
complex pattern of rejection forces that change in direction and magnitude
with time, thereby creating more variety in the oscillation patterns.
For details regarding the data analysis presented in Fig.~\ref{fig:a.Data},
see Appendix \ref{sec: Supplementary data analysis}.

\subsection{Estimation of the Vorticity-Gradient Forces - Comparison With the
Measured Accelerations}

\begin{figure*}[ht!]
\begin{centering}
\includegraphics[width=1\textwidth]{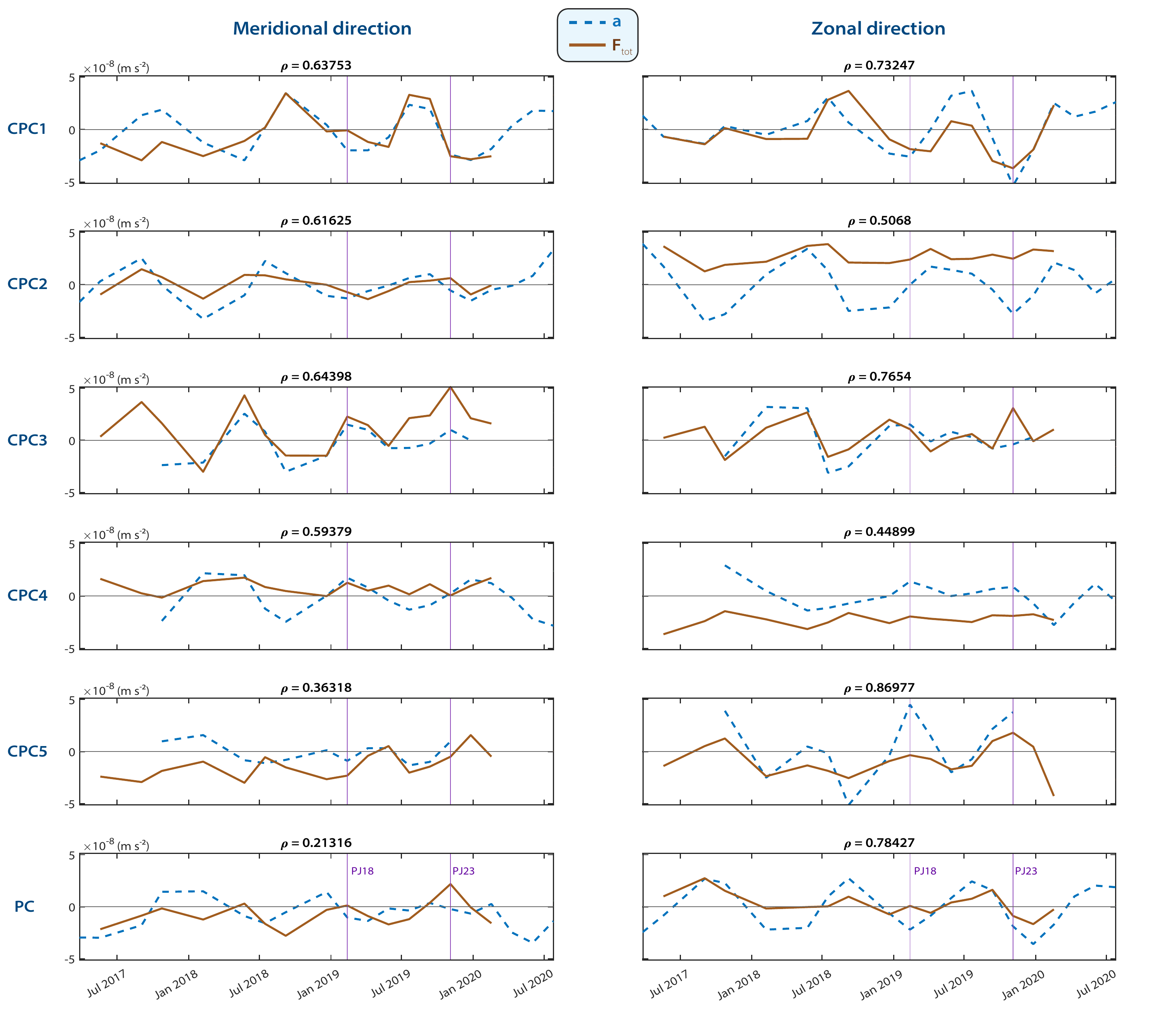}
\par\end{centering}
\caption{Comparison between the net force and the acceleration of the
cyclones, using their instantaneous location data. Each panel represents a time
series of the acceleration and force of one cyclone in one direction.
Rows represent the different cyclones. The left column is in the meridional
direction and the right column is in the zonal direction. The abscissa
represents the observation's time. The ordinate, in units of (${\rm m\,s^{-2}}$),
represents the amplitude of the two curves. The dashed blue curves
are the acceleration of the cyclones, calculated by differentiating
twice their observed positions with time. The brown lines represent
the net force on the cyclones, resulting from the sum of the mutual
cyclone rejection forces and $F_{\beta}$ (Fig.\ \ref{fig:Theory}).
The solid brown and dashed blue curves are expected to follow each
other according to the suggested force balance. The Pearson correlation
coefficient ($\rho$) between these two curves is presented above
each panel. \label{fig: F vs dudt}}
\end{figure*}
To validate that the the observed motion (Fig.~\ref{fig:a.Data})
results from the hypothesized mechanism laid out in the previous section
(Fig.\ \ref{fig:Theory}), it is possible to estimate the net force
acting on each cyclone according to its latitude and its distance
from the other cyclones at each point in time. From Newton's second
law, we expect that the acceleration ($\mathbf{a}_{i}$, where $i$
is an index representing the respective cyclone on which the forces
are calculated), estimated by differentiating the cyclone's observed position data
in time, should match 
\begin{equation}
\mathbf{a}_{i}=\underset{F_{{\rm tot,i}}}{\underbrace{\underset{j\neq i}{\sum}\mathbf{F}_{ji}+\mathbf{F}_{\beta i}}},\label{eq: acc vs force eq}
\end{equation}
where $\mathbf{F}_{ji}$ is the rejection force vector (per unit mass) by cyclone
$j$, and $\mathbf{F}_{\beta i}$ is the beta-drift force vector (per
unit mass). The forces are ``Rossby forces'' \cite{rossby1948displacements},
calculated by a concentric integration of the Coriolis force around
the cyclone, where the background vorticity alters respectively for
each force term (see Appendix \ref{sec: Supplementary data analysis}).
These calculations require determining 3 global parameters, and 3
characteristic numbers for each of the six cyclones, which dictate
their maximum velocity, their size and their morphology. These free
parameters are chosen here to stay constant with respect to time,
albeit in reality they may also have temporal variations.

To resolve these unknown parameters, an optimization procedure is
used to provide the best overall match between the two sides of Eq.\ \ref{eq: acc vs force eq}.
For the three cyclone characteristic numbers, up to $20\%$ variability
between cyclones was allowed (30\% in the case of the morphology factor), relative to mean estimated values.
Of the 3 global parameters determined by the optimization algorithm
is $K_{{\rm def}}$, which is the amount by which the forces are deflected
clockwise. This deflection accounts for the fact that after the beta-gyres
form, they can be advected by the tangential velocity of the cyclones
(clockwise in the case of cyclones in the southern hemisphere) such that the net force is
deflected, to some extent, in the clockwise direction \cite{fiorino1989some}.
In GK21, the analysis of the CPCs latitude of equilibrium assumed
a balance between the PC's rejection force and the beta-effect; here,
also taking into account the spatial distribution of the cyclones,
we find that rejection of CPCs by CPCs also has a significant component
in the meridional direction. The counter balance offered here against
this effect, is that turbulence around the pole can partially homogenize
the vorticity of the cyclones away from their centers, thereby diminishing
their respective gradients and the resulting rejection forces. To
take this into account, a factor $K_{{\rm trb}}$ is applied in the
calculation of $\mathbf{F}_{ji}$. The third global parameter is $R_{{\rm int}}$,
which determines what fraction of the cyclones' cores is being integrated
to calculate the forces. The values determined for the unknown parameters
and all other details of the calculations are elaborated in Appendix \ref{sec: Supplementary data analysis}.

The resulting forces ,calculated according to the observed instantaneous locations of the cyclones, exhibit a good match with
the respective accelerations (Fig.\ \ref{fig: F vs dudt}). This
match is particularly pronounced when looking on CPCs 1 and 3; however,
all cases achieve positive correlations ($\rho$ in Fig.\ \ref{fig: F vs dudt})
that are mostly higher than $0.5$. Some mismatch between the curves
may be a result of turbulence and impacts of cyclones which arrive
at the polar region due to beta-drift and transfer momentum to the
6 cyclones. Such impacts would only directly affect the acceleration
of the impacted cyclones, but would not have a signature in the calculation
of the forces. To highlight this, two PJs (PJ18\citep{adriani2020two},
\& PJ23\citep{mura2021oscillations}) when an intruder cyclone was observed
in the polar ring are marked by purple lines (Fig.\ \ref{fig: F vs dudt}),
but the impact's effects may only take place at a specific time during
the $\sim3.5$ months between the preceding PJ and the following PJ. In the meridional direction
of the PC, the relatively low correlation might be expected, as the
extremely low $\beta$ (and therefore $\mathbf{F}_{\beta}$) near
the pole leaves it more susceptible to noise. We note here that the acceleration time series presented in Fig.\ \ref{fig: F vs dudt}, differentiated from the position data, which is spaced $\sim 53$ days between data points, only give a measure of the long time-scale accelerations, while instantaneous accelerations can be somewhat different. This is reasonable for the comparison of the vorticity-gradient forces, which only act on long time-scales.

To reassure the reader that the implemented optimization cannot produce
this match between acceleration and force out of thin air, we stress
that all parameters are constant in time, which means that all the
temporal trends in Fig.\ \ref{fig: F vs dudt}, are direct consequences
of the observational data. This implies that without having the underlying
physics correct, it is not likely that any set of parameters would be able to produce correlations
in the time series, let alone correlate well on the 12 independent
cases (Fig.\ \ref{fig: F vs dudt}). To support this statement, a test with 300 sets of random motion is generated and optimized with the same procedure as was done for Fig.\:\ref{fig: F vs dudt}, showing very weak correlations between the random accelerations and the resulting forces (see Appendix \ref{sec: robustness analysis}, and Fig.\:\ref{fig: random correlations analysis}). Also, the sensitivity of the results (Fig.\:\ref{fig: F vs dudt}) to variations in the optimized parameters is analysed in Fig.\:\ref{fig: correlation sensitivity analysis}. Overall, the match between the forces and the accelerations supports the claim that these vorticity-gradient forces indeed drive the observed motion of the cyclones.

One approach to support the results of Fig.\ \ref{fig: F vs dudt} is to analytically estimate the oscillations period from the vorticity-gradient forces. Here, we consider a simple harmonic oscillation of one CPC and two static neighbors in the zonal direction (Fig.\:\ref{fig:freq geo}). Taking a derivative of the mean rejection force per unit mass ($F$) with a zonal perturbation ($x_0$) from the position of equilibrium gives the spring constant, from which the oscillation period ($T_{\rm N}$) is derived (see full details in Appendix \ref{sec: analytic frequency}) as 
\begin{equation}
    T_{{\rm N}}=2\pi\left(\frac{\partial F}{\partial x_{0}}\right)^{-\frac{1}{2}}.
\label{eq: Osc period}
\end{equation}
Estimating Eq.\ \ref{eq: Osc period} with the values used for Fig.\ \ref{fig: F vs dudt} gives $T_{{\rm N}}\approx15$ months, which is relatively close, given the simplicity of the model, to the observed 12-month period. The small overshoot is probably due to the higher complexity of the full 6-body problem. In addition, this analytic analysis predicts how would the oscillations change for different cyclone profiles and for different distances between cyclones (Fig.\:\ref{fig:freq res}), showing that the oscillations would only be observable for a narrow range of inter-cyclone distances. 

\subsection{Forced-Cyclones Model}

\begin{figure*}[ht!]
\begin{centering}
\includegraphics[width=0.9\textwidth]{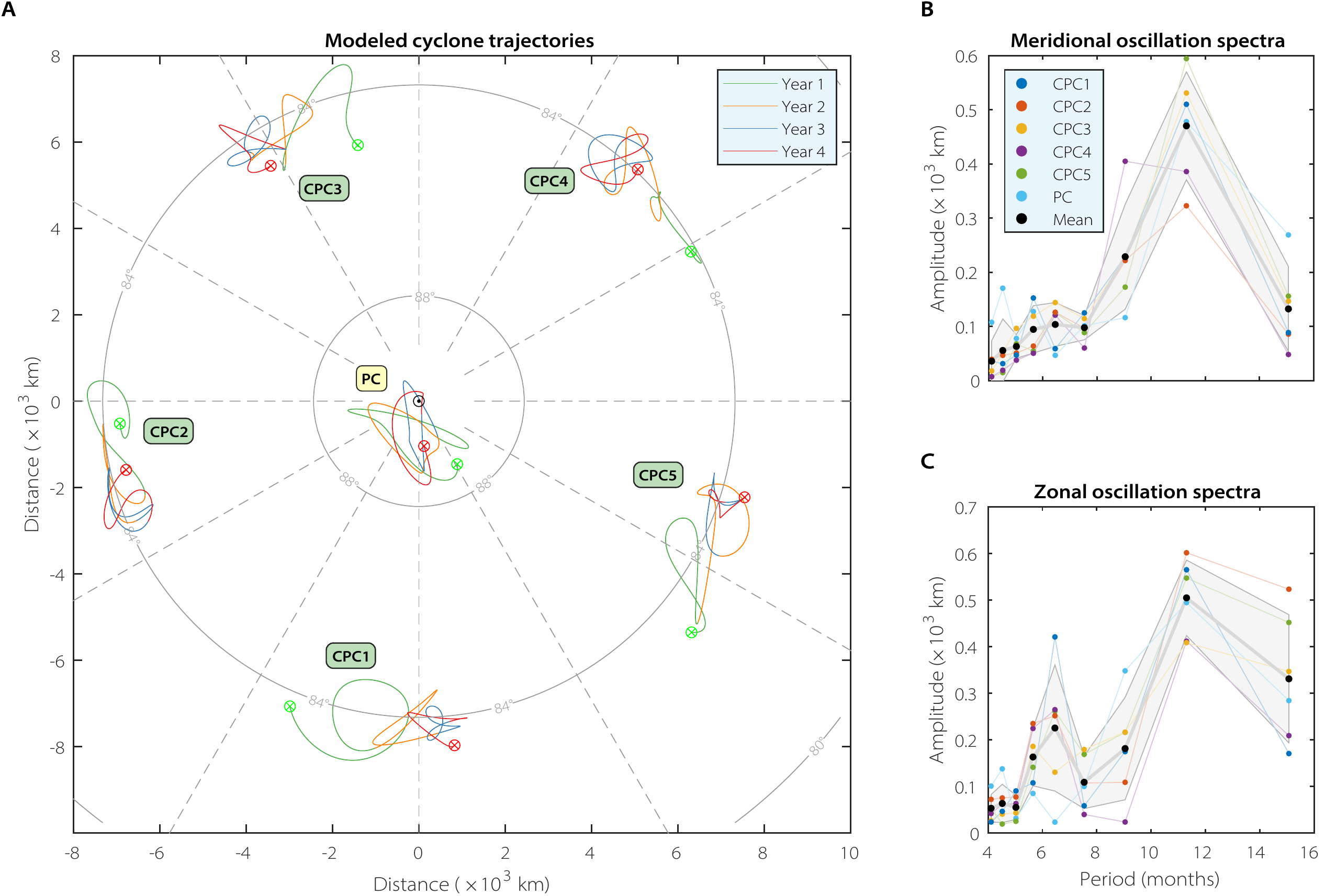}
\par\end{centering}
\caption{The modelled trajectories and spectra of Jupiter's south polar
cyclones. (a) The trajectories of the modeled cyclones,
moving in time according to Eq.~\ref{eq: acc vs force eq} and Eq.~\ref{eq:model}.
The green 'X's denote the initial position of each cyclone, corresponding
to the observed positions at PJ4. The red 'X's are the simulated final
positions at PJ30. The trends change color after each year of simulation.
An animation of the simulation is available in the SI (Movie~S2).
(b,c) Zonal and meridional modelled oscillation spectra,
respectively. The black dots and gray shades represent the mean and
the 1 STD range around it, respectively, calculated for each period
between the 6 cyclones. These spectra is presented in terms of oscillation energy in Fig.\:\ref{fig: 3 power spectra}. \label{fig:Model results}}
\end{figure*}
To further illustrate how the described mechanism (Fig.\ \ref{fig:Theory})
results in the observed motion (Fig.~\ref{fig:a.Data}), a model
that simulates the forces and the resulting trajectories of each cyclone
with time is constructed. The model advances in time according to
\begin{equation}
\begin{array}{c}
\frac{\partial\mathbf{u}_{i}}{\partial t}=\mathbf{a}_{i},\\
\frac{\partial\mathbf{x}_{i}}{\partial t}=\mathbf{u}_{i},
\end{array}\label{eq:model}
\end{equation}
where $t$ is time, $\mathbf{a_{i}}$, $\mathbf{u}_{i}$ and $\mathbf{x}_{i}$
are the acceleration (given by Eq.\ \ref{eq: acc vs force eq}),
velocity and position vectors of cyclone $i$, respectively. For initial
conditions, we use the observed locations of the cyclones at PJ4 (green
crosses in Fig.~\ref{fig:a.Data}a), and for initial velocity we
differentiate with time the locations during the first two observations.
To find the unknown parameters, as described in the previous section,
another optimization is performed, this time to get the simulated
trajectories and spectra (Fig.\;\ref{fig:Model results}) as close as possible to the observations
(Fig.\ \ref{fig:a.Data}). The final values determined for these
parameters and all other details of the model are described in Appendix \ref{sec: oscillation model}. 

As the model has no dissipation of energy, it is susceptible to resonances. However, dissipating the energy would mean that without also inserting energy, the motion would cease. To insert energy would require forcing the cyclones with artificial frequencies and amplitudes, thus biasing the model results. The optimization procedure, on the other hand, manages to avoid resonance by fine-tuning of the parameters, without insertion of additional terms to the forces balance. For this, the model
cannot use the parameter values used for Fig.\:\ref{fig: F vs dudt}, as
it results in a resonant mode. 
Alternatively, using the model set for the analysis of Fig.\:\ref{fig: F vs dudt} shows good correlations, but the magnitudes have mismatches between the forces and the accelerations. Nevertheless, the
differences between the final parameter sets of the two analyses (Fig.\:\ref{fig: F vs dudt} and Fig.\:\ref{fig:Model results}) are relatively
small, where both sets are within a reasonable physical range (see Table \ref{tab: Optimization sets}).

The model results (Fig.\ \ref{fig:Model results}) capture properly the observed motion patterns of the south polar cyclones (Fig.\ \ref{fig:a.Data}).
These patterns include both the form of their trajectories and the
motion's spectra, which peak near $12$ months with amplitudes of
$\sim400$ kilometers. These model results exemplify that the mechanism described in this study (Fig.\:\ref{fig:Theory}) can, in fact, result in the observed
motion.
The similarity between the model and the observations adds additional support that the theory outlined in
GK21 and here, can explain the transient motion of Jupiter's PCs
and CPCs, in addition to describing their equilibrium states.

\section{Discussion}

By showing that vorticity gradients, induced both by the polar cyclones
and by $\beta$, control the leading order balance of both the statistic
steady-state \cite{gavriel2021number} and the dynamic motion of the
polar cyclones (Fig.\ \ref{fig: F vs dudt}), we surmise that
Jupiter's polar flow regime is largely barotropic,
when looking at the scale of the cyclones (scales bigger than $\sim500$\ km).
This implies on the cyclones' depth, where either the thin shallow-water
equations, indicating on shallow cyclones, or deep 2D framework can
capture the cyclone's motion. If the cyclones are indeed barotropic
and deep, in agreement with the Taylor-Proudman theorem, they would
extend uninterruptedly down to the semi-conducting region, where they
would stop spinning due to magnetic drag \cite{Liu2008,Dietrich2018,duer2019,kaspi2020comparison}.
In this case, heat convection from Jupiter's core would be a plausible
energy source for the cyclones \cite{yadav2020deep,yadavbloxam2020deep,garcia2020deep,cai2021deep}. These two options should come to mind when trying to analyze
how the cyclones were formed, how they are maintained, and what determines
their strength and morphology.

The north pole of Jupiter is expected to demonstrate similar dynamics
for its cyclones as highlighted in this study for the south pole.
However, as the available infrequent data do not allow a similar detailed
analysis, this test is left for future studies, when more data might
be gathered. For the south pole, while matching the cyclone's acceleration
and position-based forces (Fig.\ \ref{fig: F vs dudt}) required
resolving unknown coefficients with an optimization procedure, the
high correlations between them were found to be a robust feature,
regardless of the determined values (within a sensible range, see Fig.\:\ref{fig: correlation sensitivity analysis}). For
the robustness of the model simulation, unoptimized runs with simple assumptions
and identical 6 cyclones also produce oscillation patterns similar
to the observations (Fig.\:\ref{fig: Ideal model run}). Overall, the results of both observational
analysis (Figs.\ \ref{fig: F vs dudt}) and of
an idealized time-evolution model (Fig.\ \ref{fig:Model results}),
establish the role of vorticity-gradient forces (Fig.\ \ref{fig:Theory})
as the driving mechanism of the Jovian south polar cyclones' observed
motion (Fig.\ \ref{fig:a.Data}), and highlight the importance of
considering these forces in future studies on gas giants' polar dynamics
and vortex motions at large.

\subsection*{Open Research}

No new data sets were generated during the current study. The data
analysed in this study were published by Mure et al. (2021)\citep{mura2021oscillations} (DOI: \url{https://doi.org/10.1029/2021GL094235}), as cited in the
text. The figure files and Matlab scripts are available at: https://doi.org/10.5281/zenodo.6611745.
%The MATLAB codes used in this paper %are available on request
%from N.G.

\subsection*{Acknowledgements}
This research has been supported by the Minerva Foundation with funding
from the Federal German Ministry for Education and Research and the
Helen Kimmel Center for Planetary Science at the Weizmann Institute
of Science.\nocite{chan1987analytical}

%\paragraph*{Author contributions}

%N.G. designed the study, performed the calculations and wrote the
%paper together with Y.K.

%\paragraph*{Correspondence and requests for materials}

%Should be addressed to N.G. (nimrod.gavriel@weizmann.ac.il)

%
\subsection*{Competing Interests}

Authors declare that they have no competing interests.

\bibliographystyle{naturemag}
\bibliography{Nimrodbib}

\renewcommand{\thefigure}{{\arabic{figure}}}
\setcounter{figure}{0}     
\renewcommand{\thetable}{S\arabic{table}} 
\renewcommand{\thefigure}{S\arabic{figure}}
\renewcommand{\theHtable}{Supplement.\thetable}
\renewcommand{\theHfigure}{Supplement.\thefigure}
\appendix
\onecolumn

\part*{Supplementary Information}

\noindent\textbf{Movie S1.} (\href{https://doi.org/10.5281/zenodo.7040219
}{Link})
An animation of the observed trajectories of the south polar cyclones, as illustrated in Fig. \ref{fig:a.Data}a, along the 45-months of observations. The blue arrows represent the estimated net forces on the cyclones, as presented in Fig. \ref{fig: F vs dudt}.
%Type or paste caption here.
%upload your movie(s) to AGU's journal submission site and select, "Supporting Information %(SI)" as the file type. Following naming convention: ms01.

\noindent\textbf{Movie S2.} (\href{https://doi.org/10.5281/zenodo.7040219
}{Link})
An animation of the simulated trajectories of the south polar cyclones, as illustrated in Fig. \ref{fig:Model results}a, along 45-months. The blue arrows represent the net forces on the cyclones.

\section{Data analysis methods} \label{sec: Supplementary data analysis}
\subsection{Trajectories and spectra}
In this subsection the procedures done for creating Fig.\:\ref{fig:a.Data} are elaborated. In this study we used the data acquired
by the Jovian Infrared Auroral Mapper (JIRAM) and published in
\cite{mura2021oscillations}. Specifically, the used data for the
observed locations of the south polar cyclones per PJ is listed in
Table~1 of \cite{mura2021oscillations}. Here, this data was converted to Cartesian coordinates ($x,y$)
via the transformation
\begin{linenomath*}
\begin{equation}
x=-R_{{\rm J}}\cos\theta\,\sin\lambda,\;\;\;y=R_{{\rm J}}\cos\theta\,\cos\lambda,
\end{equation}
\end{linenomath*}
where $R_{{\rm J}}$ is the mean radius of Jupiter ($=69,911$ km),
$\theta$ is latitude and $\lambda$ is longitude. The Cartesian data
was interpolated for the periods between PJs using a cubic ``spline''
interpolation. This interpolated data is plotted as solid lines in Fig.\:\ref{fig:a.Data},
where the dots represent the actual published data. The trajectories can also be presented as a time series of latitude and longitude (Fig.\:\ref{fig: 1 time_series}), emphasizing the dominance of the 12-month period oscillations.

\begin{figure}
\begin{centering}
\includegraphics[width=1\columnwidth]{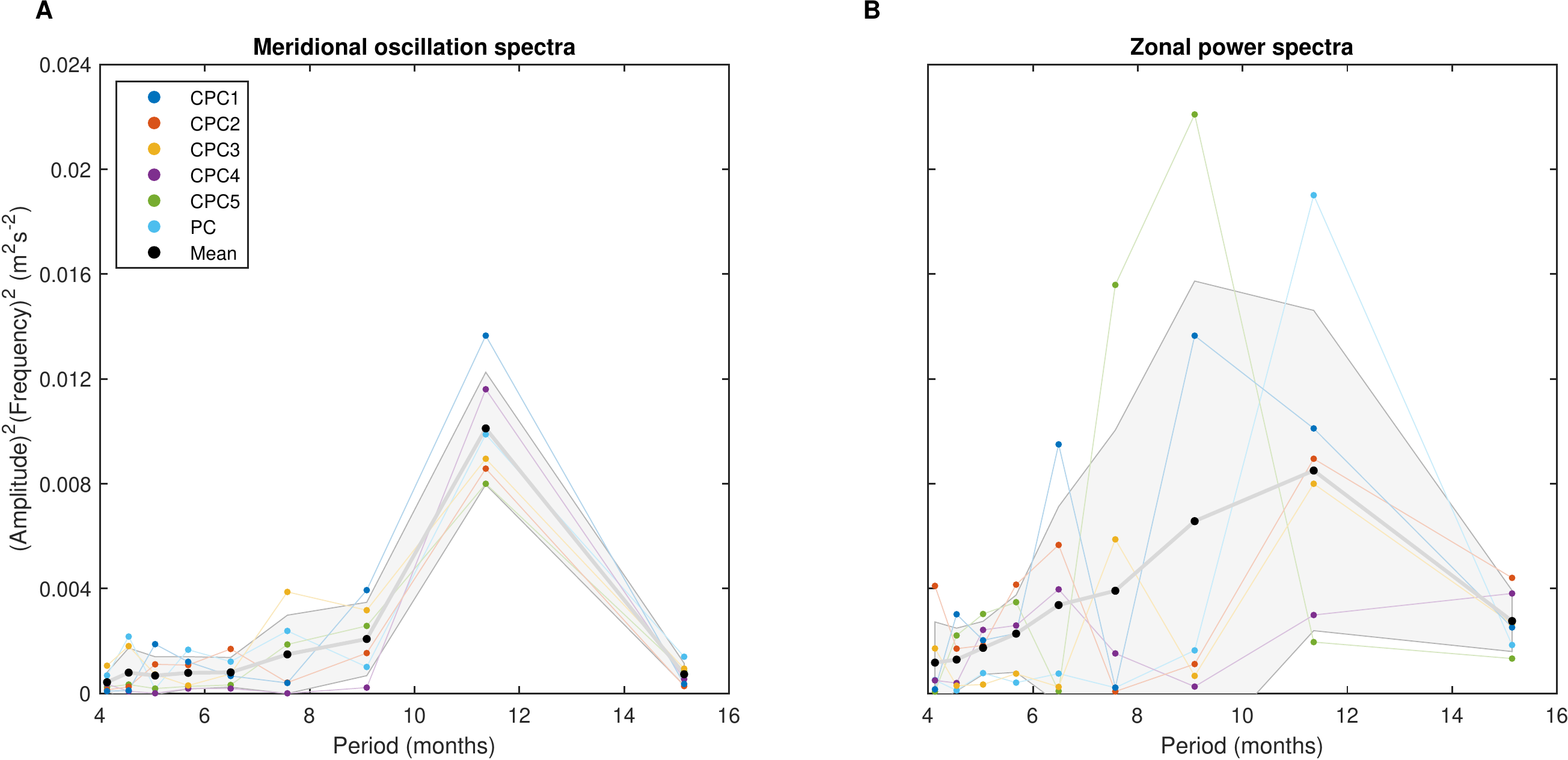}
\par\end{centering}
\caption{Energy spectra of the observed motion. Here, panels a-b are the same as Fig.\:\ref{fig:a.Data} panels b and c, but here the ordinate represents the energy of the oscillations (amplitude$^2$ frequency$^{2}$) rather than just the amplitude ($A_n$).  \label{fig: 1 power spectra}}
\end{figure}

\begin{figure}
\begin{centering}
\includegraphics[width=1\columnwidth]{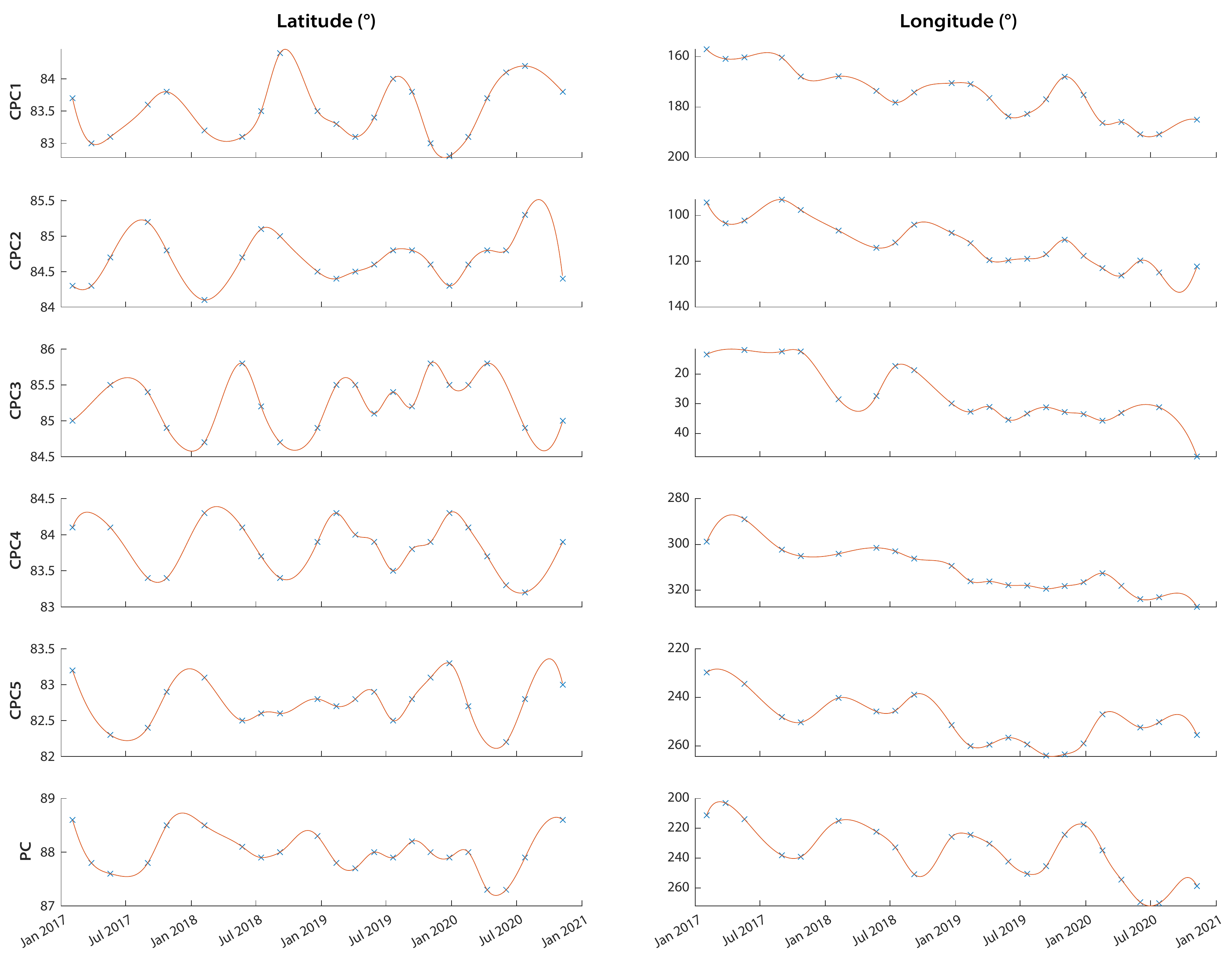}
\par\end{centering}
\caption{Time series of the observed zonal and meridional motion of Jupiter's south polar cyclones. Here, the blue 'X's represent the observed values, while the red lines represent the interpolations between the observations (the same interpolations as presented in Fig.\:\ref{fig:a.Data}). The longitude time series were previously plotted in Mura2021 (their Fig.\:2).  \label{fig: 1 time_series}}
\end{figure}

For plotting Fig. \ref{fig:a.Data}b-c, the interpolated data is
first transformed back to latitude and longitude by
\begin{linenomath*}
\begin{equation}
\theta=\cos^{-1}\left(\frac{\sqrt{x^{2}+y^{2}}}{R_{{\rm J}}}\right),\;\;\;\lambda=\tan^{-1}\left(\frac{y}{x}\right)-\frac{\pi}{2},\label{eq: cart to sph}
\end{equation}
\end{linenomath*}
to evaluate the cumulative zonal ($L_{{\lambda}}$) and meridional ($L_{{\theta}}$)
distances traveled by a cyclone $i$ at time step $k$ as
\begin{linenomath*}
\begin{equation}
\begin{array}{cc}
L_{{\lambda}(i,k)}= & L_{{\lambda}(i,k-1)}+R_{{\rm J}}\cos\left(\theta_{i,k}\right)\left(\lambda_{i,k}-\lambda_{i,k-1}\right),\\
L_{{\theta}(i,k)}= & L_{{\theta}(i,k-1)}+R_{{\rm J}}\cos\left(\theta_{i,k}-\theta_{i,k-1}\right),
\end{array}\label{eq: meridional and zonal time series}
\end{equation}
\end{linenomath*}
where we define $L_{{\lambda}(i,1)}=L_{{\theta}(i,1)}=0$. Then, after
removing the mean and linear trends, a fast Fourier transform (FFT)
was performed on the time series defined by Eq.~\ref{eq: meridional and zonal time series}.
This calculation produces the oscillation amplitudes, $A_{i,n}\left(\nicefrac{T}{n}\right)$,
where $T$ is the duration of the time series $\left(\sim45\,{\rm months}\right)$
and $n$ is a natural number. As the analyzed location data are 53
days apart, only oscillation periods larger than 4 months are displayed
(Fig. \ref{fig:a.Data}b-c), thus not regarding calculated spectra
that are artifacts of the interpolation method. Similarly, the displayed
spectra is limited to oscillation periods smaller than 16 months,
as higher periods are calculated based on less than three occurrences
in the available data set, and are thus statistically insignificant.
The black dots in Fig. \ref{fig:a.Data}b-c represent the mean (between
cyclones) spectra,
\begin{linenomath*}
\begin{equation}
\overline{A_{n}}=\frac{1}{6}\sum_{i=1}^{6}A_{i,n},
\end{equation}
\end{linenomath*}
and the black shades frame the area
\begin{linenomath*}
\begin{equation}
\overline{A_{n}}\pm\sqrt{{\rm Var}\left(A_{1,n},A_{2,n},A_{3,n},A_{4,n},A_{5,n},A_{6,n}\right)},
\end{equation}
\end{linenomath*}
representing the standard error around the mean. In Fig.\:\ref{fig: 1 power spectra}, the spectra is presented in terms of energy instead of metric amplitudes.

\subsection{Estimating the forces acting on the cyclones and their acceleration}

To illustrate how the forces illustrated in Fig.\:\ref{fig:Theory}
result in the observed oscillatory motion (Fig.\:\ref{fig:a.Data}),
these forces are estimated according to the following layout.

\subsubsection{Velocity profile for each cyclone}

The ideal tangential velocity profile ($U_{i}\left(r_{i}\right)$,
where $r_{i}$ is the distance from the cyclone's center) around a
cyclone $i$, where $i$ is the cyclone's identifier (ranging from
1 to 6 in the SH), used in this study is \citep{chan1987analytical}
\begin{linenomath*}
\begin{equation}
U_{i}\left(r_{i}\right)=V_{i}\left(\frac{r_{i}}{R_{i}}\right)\exp\left[\frac{1}{b_{i}}\left(1-\left(\frac{r_{i}}{R_{i}}\right)^{b_{i}}\right)\right],\label{eq: Tangential velocity profile}
\end{equation}
\end{linenomath*}
where $V_{i}$, $R_{i}$ and $b_{i}$ are the maximum velocity, the
radius of maximum velocity and a steepness constant of cyclone $i$,
respectively. This ideal profile captures the observed trends of an inner
solid-rotating core and an exponential decay outside, while the connection
between the two regions is continuous. The distance, $r_{i}$, from
cyclone $i$, and the angle $\phi_{i}$ are calculated as 

\begin{linenomath*}
\begin{equation}
\begin{array}{cc}
r_{i}\left(x,y,t\right)= & \sqrt{\left(x-x_{i}\left(t\right)\right)^{2}+\left(y-y_{i}\left(t\right)\right)^{2}},\\
\phi_{i}\left(x,y,t\right)= & \tan^{-1}\left(\frac{y-y_{i}\left(t\right)}{x-x_{i}\left(t\right)}\right),
\end{array}
\end{equation}
\end{linenomath*}
where $x_{i}\left(t\right)$ and $y_{i}\left(t\right)$, are the instantaneous
coordinates of the center of cyclone $i$ at time $t$ in the $x$
and $y$ directions. The Cartesian velocities in the $x$ and $y$
directions are thus
\begin{linenomath*}
\begin{equation}
\begin{array}{cc}
u_{i}\left(r_{i},\phi_{i}\right)= & -U_{i}\left(r_{i}\right)\sin\left(\phi_{i}\right),\,{\rm and}\\
v_{i}\left(r_{i},\phi_{i}\right)= & U_{i}\left(r_{i}\right)\cos\left(\phi_{i}\right),
\end{array}\label{eq: Cartesian velocity profiles}
\end{equation}
\end{linenomath*}
respectively. 

\subsubsection{Potential vorticity}

Using Eqs.~\ref{eq: Tangential velocity profile}-\ref{eq: Cartesian velocity profiles},
the relative vorticity field due to the velocity profile of cyclone
$i$, calculated as $\xi_{i}=\frac{1}{r_{i}}\frac{\partial}{\partial r_{i}}\left(r_{i}U_{i}\right)$,
is
\begin{linenomath*}
\begin{equation}
\xi_{i}\left(r_{i}\right)=\frac{V_{i}}{R_{i}}\left(2-\left(\frac{r_{i}}{R_{i}}\right)^{b_{i}}\right)\exp\left[\frac{1}{b_{i}}\left(1-\left(\frac{r_{i}}{R_{i}}\right)^{b_{i}}\right)\right].
\end{equation}
\end{linenomath*}
The planetary vorticity is 
\begin{linenomath*}
\begin{equation}
f=2\Omega\cos\left(\frac{\sqrt{x^{2}+y^{2}}}{R_{{\rm J}}}\right),
\end{equation}
\end{linenomath*}
where $\Omega$ and $R_{{\rm J}}$ are the rotation rate and mean
radius of Jupiter, respectively.

\subsubsection{Calculating the forces}

The individual forces are calculated as the net mean integrated Coriolis
forces (sometimes referred to as a 'Rossby force', \cite{rossby1948displacements})
resulting from the interaction of the tangential velocity of the forced
cyclone with either $f$, representing the beta-drift force, or the
relative vorticity of the other cyclones, representing the mutual
rejection forces between the cyclones. The forces would be calculated
in two parts. First, the forces (per unit mass) in the $x$ and $y$
directions are integrated as 
\begin{linenomath*}
\begin{equation}
\begin{array}{ccccc}
\widetilde{F}_{\beta i,x}= & \left(\varoiintop dS_{i}\right)^{-1}\varoiintop fv_{i}dS_{i}, &  & \widetilde{F}_{\beta i,y}= & -\left(\varoiintop dS_{i}\right)^{-1}\varoiintop fu_{i}dS_{i},\\
\widetilde{F}_{ji,x}= & \left(\varoiintop dS_{i}\right)^{-1}\varoiintop K_{{\rm trb}}\xi_{j}v_{i}dS_{i}, &  & \widetilde{F}_{ji,y}= & -\left(\varoiintop dS_{i}\right)^{-1}\varoiintop K_{{\rm trb}}\xi_{j}u_{i}dS_{i},
\end{array}\label{eq: Integral for calculating the forces}
\end{equation}
\end{linenomath*}
where $\xi_{j}$ is evaluated at $r_{j}$ for cyclones $j\neq i$,
and $K_{{\rm trb}}$ is a global constant that parameterizes turbulence
in the atmosphere, which can mix and diminish the PV of each cyclone,
when integrated far from its core, around a neighbor forced cyclone.
The integrated area $dS_{i}=r_{i}dr_{i}d\phi_{i}$, is in the range
$\phi_{i}=\text{\ensuremath{\left[0,2\pi\right]}}$ and $r_{i}=\text{\ensuremath{\left[0,R_{{\rm int}}R_{i}\right]}}$,
where $R_{{\rm int}}$ is a global constant that defines what fraction
of the cyclone's core is numerically integrated.

Advection of vorticity anomalies with the tangential velocity of a
cyclone can result in a westward component to the planetary beta-drift
force \citep{fiorino1989some,chan2005physics}. Therefore, a clockwise
(in the direction of the rotation of SH cyclones) deflection of the
forces is allowed. This results in the forces having the final form
\begin{linenomath*}
\begin{equation}
\begin{array}{ccccc}
F_{\beta i,x}= & \widetilde{F}_{\beta i,x}\left(1-K_{{\rm def}}\right)+\widetilde{F}_{\beta i,y}K_{{\rm def}}, &  & F_{\beta i,y}= & \widetilde{F}_{\beta i,y}\left(1-K_{{\rm def}}\right)-\widetilde{F}_{\beta i,x}K_{{\rm def}},\\
F_{ji,x}= & \widetilde{F}_{ji,x}\left(1-K_{{\rm def}}\right)+\widetilde{F}_{ji,y}K_{{\rm def}}, &  & F_{ji,y}= & \widetilde{F}_{ji,y}\left(1-K_{{\rm def}}\right)-\widetilde{F}_{ji,x}K_{{\rm def}},
\end{array}\label{eq: Forces in x and y - after deflection}
\end{equation}
\end{linenomath*}
where $0\leq K_{{\rm def}}\leq1$ is a global constant determining
the magnitude of the clockwise deflection. The net force vector is
finally 
\begin{linenomath*}
\begin{equation}
\mathbf{F}_{{\rm tot},i}=\mathbf{F}_{\beta i}+\sum_{j\neq i}\mathbf{F}_{ji},\label{eq: Net force}
\end{equation}
\end{linenomath*}
where the $x$ and $y$ components of the force vectors are defined
by Eqs.~\ref{eq: Integral for calculating the forces}-\ref{eq: Forces in x and y - after deflection}.

\subsubsection{Finding the unknown parameters}

The theory outlined in this manuscript suggests that $\mathbf{F}_{{\rm tot},i}$,
an idealized parameterization for the force density acting on each
cyclone according to the latitude of the cyclone ($\mathbf{F}_{\beta,i}$)
and the distance to all the other cyclones ($\mathbf{F}_{ji}$), should
be equal to the acceleration of the same cyclone, according to Newton's
second law. The acceleration, $\mathbf{a}_{i}$, is estimated from
the position data using finite difference derivative approximations
in the $x$ and $y$ directions on the observed locations time-series. To estimate $\mathbf{F}_{{\rm tot},i}$,
unknown parameters should be determined. These parameters include
three global constants: $K_{{\rm def}}$, $K_{{\rm trb}}$ and $R_{{\rm int}}$,
and 3 sets of individual constants for the six cyclones: $V_{i}$,
$R_{i}$ and $b_{i}$, which, for simplicity, are assumed here to
remain constant during the 4-year time series. To determine these
parameters, an optimization procedure is executed, which searches
for a set that minimizes a cost function containing 4 additive parts of equal weight: (i) 1 minus the mean correlation between $\mathbf{F}_{{\rm tot},i}$ and $\mathbf{a}_{i}$; (ii) the    differences between the time series of
$\mathbf{F}_{{\rm tot},i}$ and $\mathbf{a}_{i}$, averaged for all 6 cyclones; (iii) The difference between the maximum positive amplitudes of $\mathbf{F}_{{\rm tot},i}$ and $\mathbf{a}_{i}$; (iv) The difference between the maximum negative amplitudes of $\mathbf{F}_{{\rm tot},i}$ and $\mathbf{a}_{i}$. The optimization was executed using Matlab's fmincon function, which uses the 'interior-point' algorithm.
The resulting set, used for plotting Fig.\:\ref{fig: F vs dudt},
is displayed in Table.~\ref{tab: Optimization sets} under ``Set
1''.
%
 %\begin{table}
% \caption{Time of the Transition Between Phase 1 and Phase 2$^{a}$}
% \centering
 %\begin{tabular}{l c}
% \hline
 % Run  & Time (min)  \\
% \hline
 %  $l1$  & 260   \\
%   $l2$  & 300   \\
%   $l3$  & 340   \\
%   $h1$  & 270   \\
%   $h2$  & 250   \\
%   $h3$  & 380   \\
%   $r1$  & 370   \\
%   $r2$  & 390   \\
% \hline
% \multicolumn{2}{l}{$^{a}$Footnote text here.}
% \end{tabular}
% \end{table}
\begin{table}
\caption{Parameter sets determined for Fig.\:\ref{fig: F vs dudt}
(Set\ 1), Fig.\:\ref{fig:Model results} (Set\ 2), and the mean
and variance between them. The variances of the per-cyclone parameters
($V_{i},R_{i},b_{i}$) are between the combined 12 numbers of both
sets. $\rho_{\rm{mean}}$ is the correlation between the observed  acceleration and the net force calculated with the respective set, averaged for the 6 cyclones and the 2 directions. Set 3 is the result of optimization with only 2 variables (Fig.\:\ref{fig: F vs a 2vars}). The ideal case lists the values determined for the model robustness
tests (Section S3)\\\label{tab: Optimization sets}}
\centering
\begin{tabular}{|c||c|c|c|c|c|c||c|}
\hline 
 & {\footnotesize{}$K_{{\rm def}}$} & {\footnotesize{}$K_{{\rm trb}}$} & {\footnotesize{}$R_{{\rm int}}$} & {\footnotesize{}$\begin{array}{c}
V_{1}\\
V_{2}\\
V_{3}\\
V_{4}\\
V_{5}\\
V_{6}
\end{array}$(${\rm ms}^{-1}$)} & {\footnotesize{}$\begin{array}{c}
R_{1}\\
R_{2}\\
R_{3}\\
R_{4}\\
R_{5}\\
R_{6}
\end{array}$(km)} & {\footnotesize{}$\begin{array}{c}
b_{1}\\
b_{2}\\
b_{3}\\
b_{4}\\
b_{5}\\
b_{6}
\end{array}$}
 & {\footnotesize{}$\rho_{\rm{mean}}$}
\\
\hline 
\hline 
{\footnotesize{}Set~1} & {\footnotesize{}$0.1889$} & {\footnotesize{}$0.3176$} & {\footnotesize{}$0.0643$} & {\footnotesize{}$\begin{array}{c}
92.7\\
78.01\\
86.19\\
87.17\\
79.17\\
82.22
\end{array}$} & {\footnotesize{}$\begin{array}{c}
1,192\\
1,031\\
1,115\\
1,031\\
1,060\\
1,069
\end{array}$} & {\footnotesize{}$\begin{array}{c}
0.678\\
0.893\\
0.748\\
1.028\\
0.632\\
0.817
\end{array}$} & {\footnotesize{}$0.598$}\\
\hline 
{\footnotesize{}Set~2} & {\footnotesize{}$0.1148$} & {\footnotesize{}$0.3128$} & {\footnotesize{}$0.1041$} & {\footnotesize{}$\begin{array}{c}
79.68\\
86.11\\
89.34\\
85.63\\
81.74\\
84.98
\end{array}$} & {\footnotesize{}$\begin{array}{c}
1,042\\
1,134\\
1,112\\
1,093\\
1,051\\
1,145
\end{array}$} & {\footnotesize{}$\begin{array}{c}
0.702\\
0.743\\
0.799\\
0.821\\
0.832\\
0.730
\end{array}$}& {\footnotesize{}$0.622$}

\\
\hline 
{\footnotesize{}Mean} & {\footnotesize{}$0.1519$} & {\footnotesize{}$0.3152$} & {\footnotesize{}$0.0842$} & {\footnotesize{}$84.41$} & {\footnotesize{}$1,090$} & {\footnotesize{}$0.7852$} &
\\
\hline 
{\footnotesize{}STD} & {\footnotesize{}$0.0524$} & {\footnotesize{}$0.0034$} & {\footnotesize{}$0.0281$} & {\footnotesize{}$4.3816$} & {\footnotesize{}$51$} & {\footnotesize{}$0.1059$} &
\\
\hline 
{\footnotesize{}STD ($\%$)} & {\footnotesize{}$34.5\%$} & {\footnotesize{}$1.1\%$} & {\footnotesize{}$33.4\%$} & {\footnotesize{}$5.2\%$} & {\footnotesize{}$4.7\%$} & {\footnotesize{}$13.5\%$} &
\\
\hline 
\hline 
{\footnotesize{}Set 3} & {\footnotesize{}$0$} & {\footnotesize{}$0.2768$} & {\footnotesize{}$0.0472$} & {\footnotesize{}$85$} & {\footnotesize{}$1,100$} & {\footnotesize{}$0.78$} &
{\footnotesize{}$0.536$}

\\
\hline 
\hline 
{\footnotesize{}Ideal} & {\footnotesize{}$0$} & {\footnotesize{}$1$} & {\footnotesize{}$0.1$} & {\footnotesize{}$85$} & {\footnotesize{}$1,100$} & {\footnotesize{}$0.78$} & {\footnotesize{}$0.4179$}
\\
\hline 
\end{tabular}
\end{table}

\section{Robustness and sensitivity of the correlation analysis to optimization} \label{sec: robustness analysis}
\begin{figure}
\begin{centering}
\includegraphics[width=0.7\columnwidth]{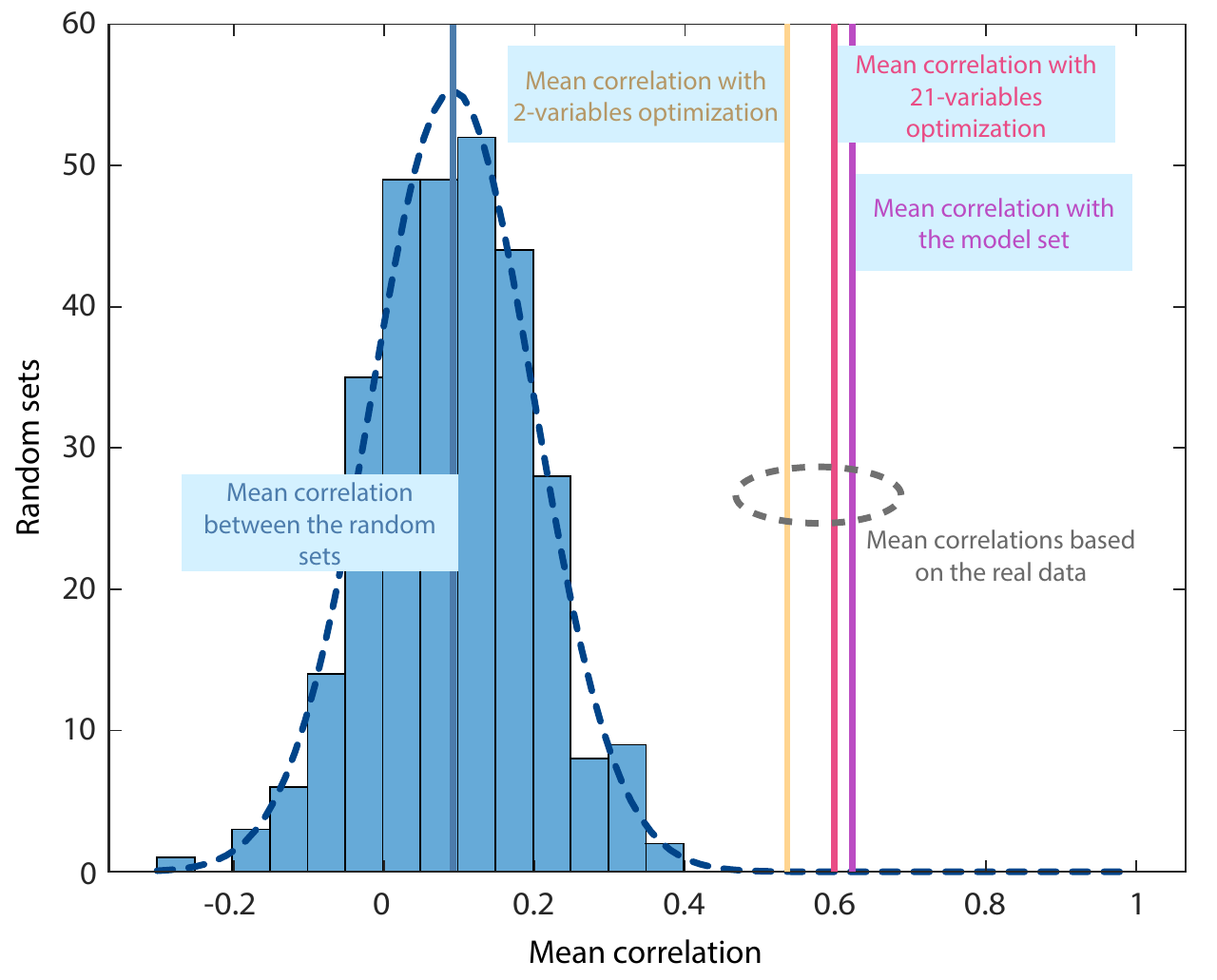}
\par\end{centering}
\caption{Histogram of the random acceleration sets results. The histogram presents the mean correlations (abscissa) where the total amount of sets is 300. The vertical blue line represents the mean of the 300 mean correlations. The dashed line represents the Gaussian distribution of the mean correlations between the 300 sets. The vertical red line represents the mean correlation of Set 1 (Tab.\;\ref{tab: Optimization sets}). The yellow vertical line represents the mean correlation of Set 3, found by optimizing only 2 variables. The purple vertical line represents the mean correlation of Set 2, which optimizes the model results (Fig.\;\ref{fig:Model results}).  \label{fig: random correlations analysis}}
\end{figure}
In this section we verify that the optimization procedure cannot find the same level of correlation in random data as it can find in the real data. Afterwards, we check the sensitivity of the results to the chosen values determined by the optimization. We begin this analysis by generating 300 sets of random data. Each set is composed of a random trajectory time series (with the total period, $\sim 4$ years, of the real data) for each of the six cyclones. The initial positions for each random set are the positions of the six cyclones at PJ4. As the vorticity-gradient forces (Eq.\;\ref{eq: Net force}) are functions of the cyclones' positions, the way to verify that the optimization cannot form correlation just by fitting parameters is to generate random forces which are independent of the cyclones' positions. For this reason, the random motion is generated by randomly forcing each cyclone. This forcing is done by accelerating each cyclone with a random magnitude between $-0.0017$ and $0.0017\; ({\rm m\:s}^{-2})$ in each direction ($x$ and $y$) for consecutive periods of $\sim 7$ days.

Each random set, having both acceleration and position time series, goes through the same optimization procedure as presented in the previous section. Then, for each optimized set, the mean correlation is calculated by averaging the 12 correlations between the forces and accelerations of the six cyclones in the zonal and meridional directions. The resulting mean correlations of the 300 random data-sets are presented as a histogram in Fig.\;\ref{fig: random correlations analysis}. It can be seen that the ability of the optimization to create a correlation by fitting the free variables is evaluated to be approximately $0.1$, whereby using the real data, it can reach $\sim 0.6$. The STD representing the Gaussian distribution of the 300 sets (Fig.\;\ref{fig: random correlations analysis}) is $\sim 0.1$, such that the probability that the mean correlation of Set 1 is a part of the random distribution is $ 0.0005\% $.

\begin{figure}
\begin{centering}
\includegraphics[width=1\columnwidth]{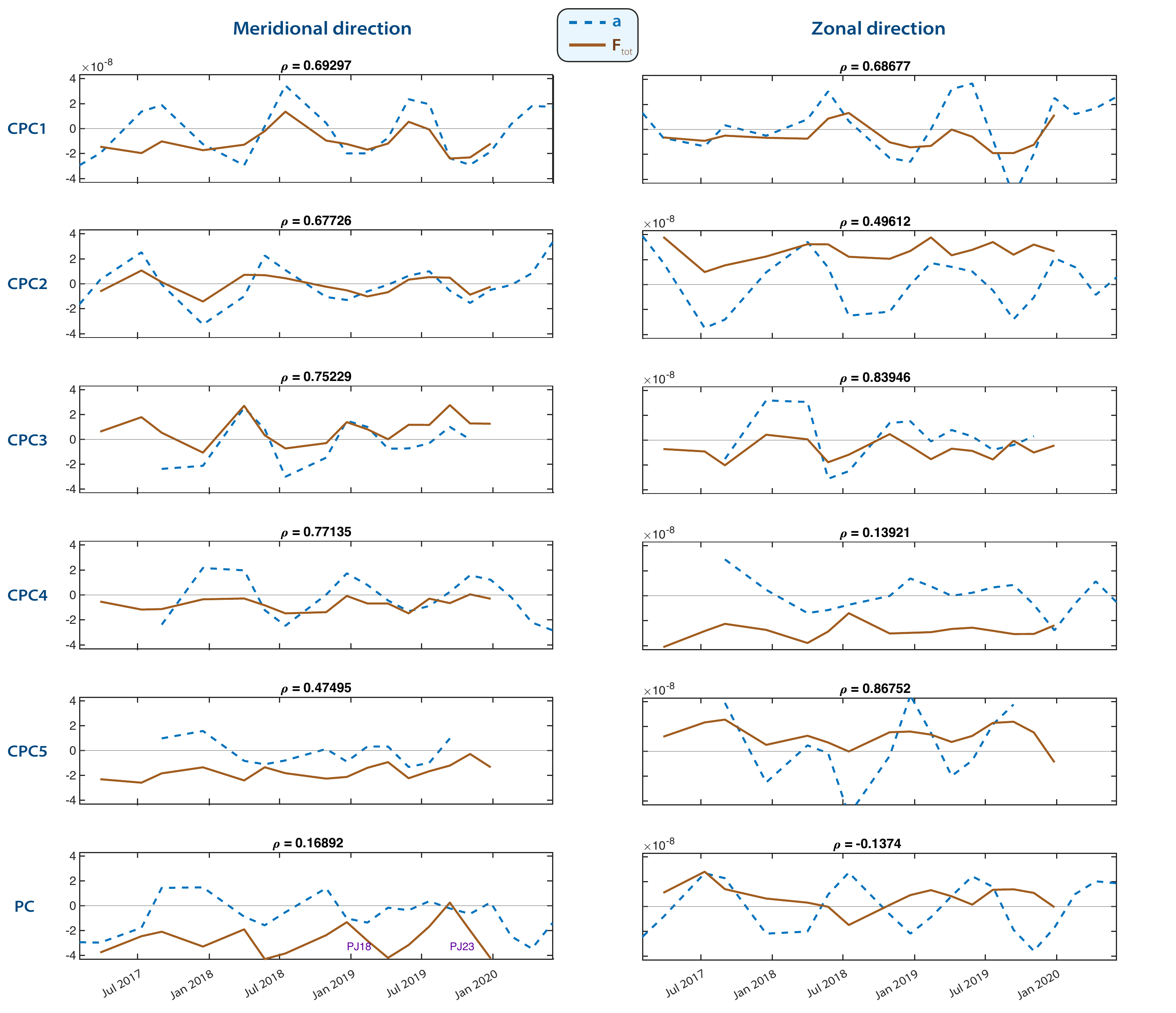}
\par\end{centering}
\caption{The same as Fig.\:\ref{fig: F vs dudt}, using Set 3 instead of Set 1 (Tab.\:\ref{tab: Optimization sets}). Here the optimization is only performed for 2 variables, where the rest of the variables are pre-defined. \label{fig: F vs a 2vars}}
\end{figure}

\begin{figure}
\begin{centering}
\includegraphics[width=1\columnwidth]{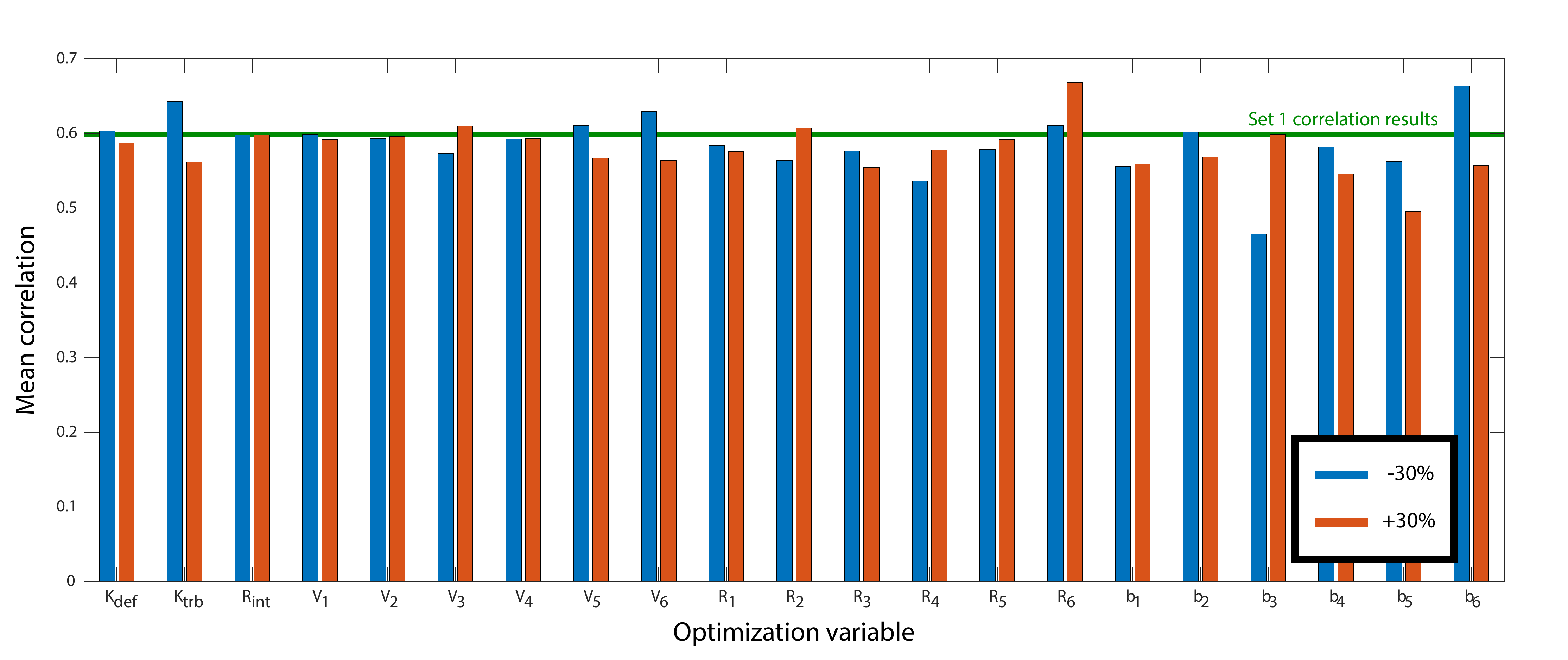}
\par\end{centering}
\caption{Sensitivity analysis for the correlation results of Fig.\:\ref{fig: F vs dudt}. Here, each optimized variable (Tab.\:\ref{tab: Optimization sets}) is decreased (blue) or increased (red) by 30\%. The ordinate is the mean correlation for every test, while the horizontal green line represents the mean correlation of the optimized set.\label{fig: correlation sensitivity analysis}   }
\end{figure}

For appreciating the sensitivity of the Fig.\:\ref{fig: F vs dudt} results to changes in the chosen values for the free parameters, we start with a test optimization run that has only 2 free parameters to optimize, while the other variables are pre-defined. In this test, all the cyclones are defined with the same size, velocity, and shape. The calculated forces resulting from this optimization still show a fine correlation with the acceleration data (Fig.\:\ref{fig: F vs a 2vars} and yellow line in Fig.\:\ref{fig: random correlations analysis}), although not as good as when the optimization had 21 free variables. In addition, a sensitivity analysis is done, where Set 1 was used as a reference for the mean correlation, and each variable is independently increased or decreased by $30\%$ from its respective value in Set 1, while the other variables have the value from Set 1. The results of this analysis (Fig.\:\ref{fig: correlation sensitivity analysis}) show relatively small sensitivities, as captured by the mean correlation, to changes in any one variable. The reason why some cases achieve a mean correlation higher than that of Set 1, is that the optimization does not only maximize correlation, but also tries to achieve a good match in the signs and magnitudes between the acceleration data and the forces.

\section{Analytic solution for the oscillations period} \label{sec: analytic frequency}
% Preview source code for paragraph 1
\begin{figure}
\begin{centering}
\includegraphics[width=0.7\columnwidth]{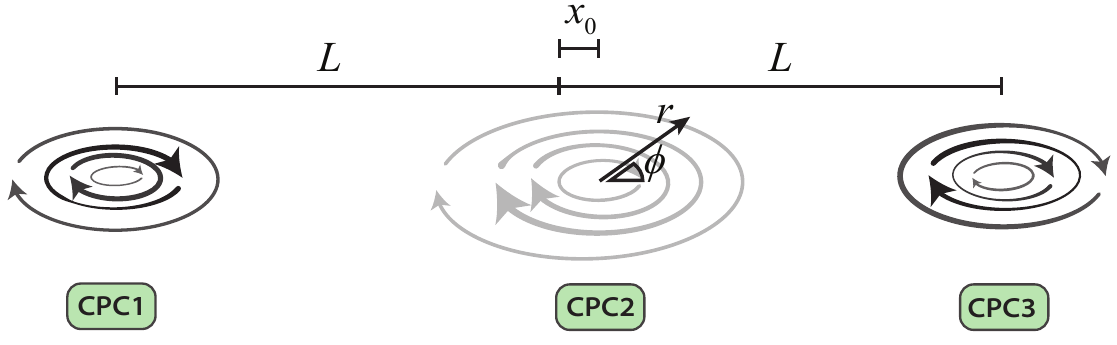}
\par\end{centering}
\caption{Schematics of the considered idealized oscillation problem. CPC1 and CPC3 are statically positioned. CPC2 is displaced by $x_{0}$
from the center between the CPCs 1 and 3, which are $2L$ apart. The polar coordinates (the radius $r$ and angle $\phi$) are defined from the center of CPC2. \label{fig:freq geo}}
\end{figure}
For finding an analytic estimate of the natural oscillation frequency,
the layout of Fig.\:\ref{fig:freq geo} is considered. The analysis
is conducted on the motion of CPC2, displaced by a small $x_{0}$
from the midpoint between the adjacent CPCs 1 and 3. The neighbor CPC1 and CPC3
are assumed to be static for simplicity. Also, the cyclones are assumed
to be identical. In a simple harmonic motion, the natural frequency
of an oscillation pattern can be calculated by 
\begin{linenomath*}
\begin{equation}
\omega_{{\rm N}}=\sqrt{\frac{k}{M}},
\end{equation}
\end{linenomath*}
where $k$ is the spring constant and $M$ is the mass of the cyclone.
The spring constant can be calculated by a perturbation from the point
of equilibrium $(x_{0}=0)$ as
\begin{linenomath*}
\begin{equation}
\frac{k}{M}=-\frac{\partial F}{\partial x_{0}},\label{eq:spring1}
\end{equation}
\end{linenomath*}
where $F$ is the mean zonal force (per unit mass) on CPC2 by the
presence of CPCs 1 and 3, calculated as the integrated Coriolis force
around CPC2 (for an asymptotic derivation of this force, see the Methods section in GK21\nocite{gavriel2021number}), 
\begin{linenomath*}
\begin{equation}
F=\left(\varoiintop dS\right)^{-1}\varoiintop K\left(\xi_{1}+\xi_{3}\right)v_{2}dS,
\end{equation}
\end{linenomath*}
where $\xi_{1}$ and $\xi_{3}$ are the vorticity contributions from
CPCs 1 and 3, respectively, $K\equiv K_{{\rm trb}}\left(1-K_{{\rm def}}\right)$ 
is a constant parameterizing the effects of turbulence and meridional deflection, and $v_{2}$ is the meridional
velocity field of CPC2. As the integration area ($dS=rdrd\phi$) is
assumed to be within the radius of maximum velocity of CPC2, we may
make the solid-rotating body approximation
\begin{equation}
v_{2}=\frac{r}{R}V\cos\phi,
\end{equation}
\begin{figure}
\begin{centering}
\includegraphics[width=0.6\columnwidth]{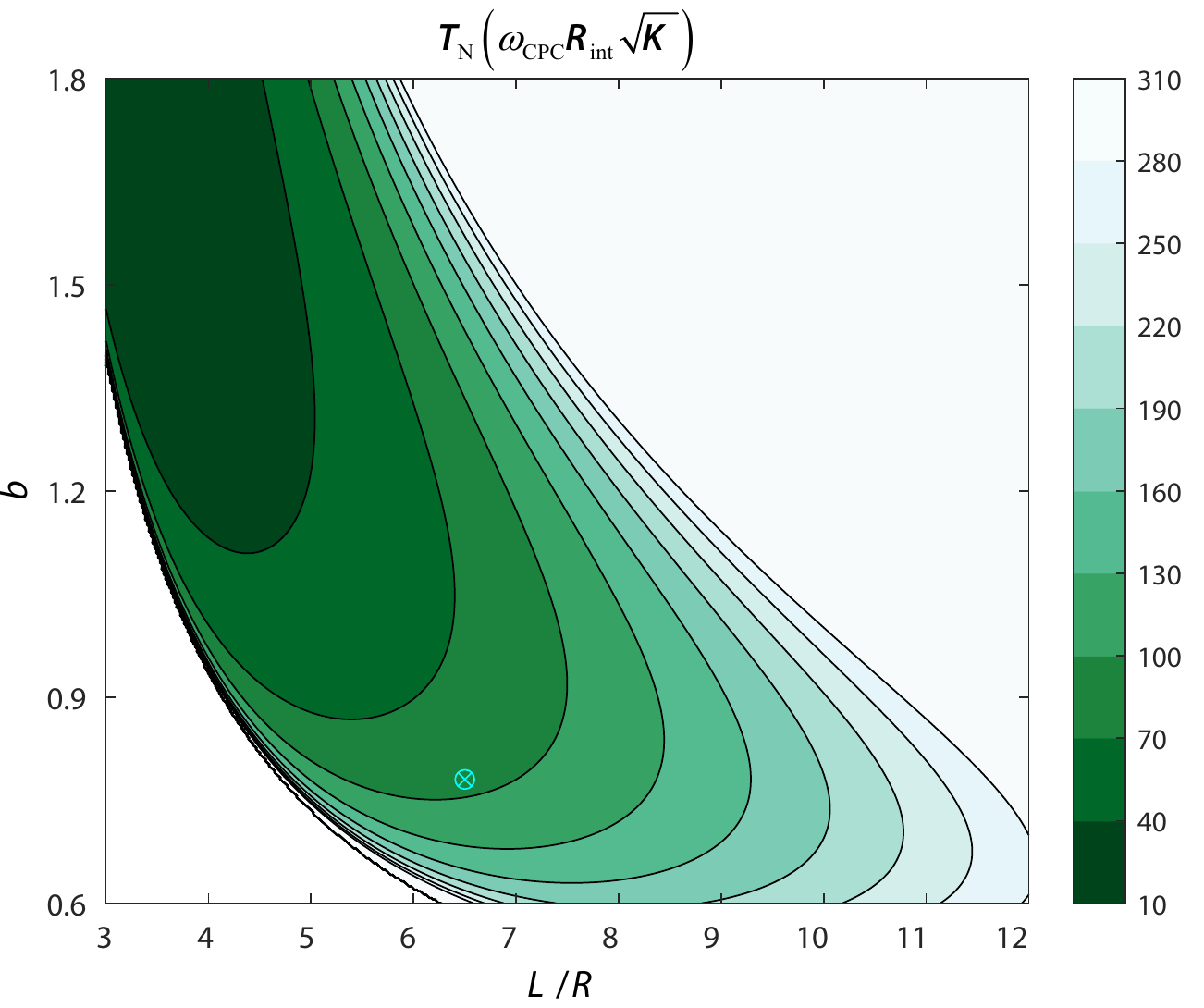}
\par\end{centering}
\caption{A contour for the values of $G=T_{{\rm N}}\left(\omega_{{\rm CPC}}R_{{\rm int}}\sqrt{K}\right)$
. The cyan 'X' is a representative value for the south pole Jovian
CPCs.\label{fig:freq res}}
\end{figure}
where $r$ and $\phi$ represent polar coordinates around the center
of CPC2, $V$ is the maximum velocity and $R$ is the radius of maximum
velocity for the CPCs (see GK21, Extended Data Figs 1-3 for reference). The vorticity field for a CPC is estimated
(see section S1) as
\begin{linenomath*}
\begin{equation}
\xi_{i}=\frac{V}{R}\left(2-\left(\frac{r_{i}}{R}\right)^{b}\right)\exp\left[\frac{1}{b}\left(1-\left(\frac{r_{i}}{R}\right)^{b}\right)\right],
\end{equation}
\end{linenomath*}
where $r_{i}$ is the distance from the respective ($i$) CPC's core,
specifically 
\begin{linenomath*}
\begin{equation}
\begin{array}{c}
r_{1}=\sqrt{\left(L+\left(x_{0}+r\cos\phi\right)\right)^{2}+\left(r\sin\phi\right)^{2}}\\
r_{3}=\sqrt{\left(L-\left(x_{0}+r\cos\phi\right)\right)^{2}+\left(r\sin\phi\right)^{2}}
\end{array}.\label{eq:r1 and r3 full}
\end{equation}
\end{linenomath*}
As $L\gg\left(x_{0}+r\cos\phi;r\sin\phi\right)$, Eq.~\ref{eq:r1 and r3 full}
can be approximated by 
\begin{linenomath*}
\begin{equation}
\begin{array}{c}
r_{1}=\sqrt{L^{2}+2L\left(x_{0}+r\cos\phi\right)}\\
r_{3}=\sqrt{L^{2}-2L\left(x_{0}+r\cos\phi\right)}
\end{array}.
\end{equation}
\end{linenomath*}
As the velocity $v_{2}$ is independent of $x_{0}$, Eq.~\ref{eq:spring1}
becomes
\begin{linenomath*}
\begin{equation}
\frac{k}{M}=-\left(\varoiintop dS\right)^{-1}\varoiintop K\left(\frac{\partial\xi_{1}}{\partial x_{0}}+\frac{\partial\xi_{3}}{\partial x_{0}}\right)v_{2}dS.\label{eq:spring2}
\end{equation}
\end{linenomath*}
Using a Taylor series expansion of order $O(x_{0}^{2})$ for $\xi_{i}$,
we get
\begin{linenomath*}
\begin{equation}
\left(\frac{\partial\xi_{1}}{\partial x_{0}}+\frac{\partial\xi_{3}}{\partial x_{0}}\right)v_{2}\approx-e^{\frac{1}{b}\left(1-\left(\frac{L}{R}\right)^{b}\right)}\frac{2\omega_{{\rm CPC}}^{2}}{L^{2}}\left(\frac{L}{R}\right)^{b}\left(b^{2}-3b\left(\frac{L}{R}\right)^{b}+\left(\frac{L}{R}\right)^{2b}-4\right)\left(r\cos\ensuremath{\phi}+x_{0}\right)r\cos\ensuremath{\phi},
\end{equation}
\end{linenomath*}
where $\omega_{{\rm CPC}}=\frac{V}{R}$. Then, integrating in the
range $\phi=\text{\ensuremath{\left[0,2\pi\right]}}$ and $r_{i}=\text{\ensuremath{\left[0,R_{{\rm int}}R_{i}\right]}}$,
Eq.~\ref{eq:spring2} becomes
\begin{linenomath*}
\begin{equation}
\frac{k}{M}=\frac{1}{2}\omega_{{\rm CPC}}^{2}R_{{\rm int}}^{2}Ke^{\frac{1}{b}\left(1-\hat{L}^{b}\right)}\hat{L}^{b-2}\left(b^{2}-3b\hat{L}^{b}+\hat{L}^{2b}-4\right),
\end{equation}
\end{linenomath*}
where $\hat{L}=RL$, is a normalized $L$. The oscillation period
is thus
\begin{linenomath*}
\begin{equation}
T_{{\rm N}}=\frac{2\pi}{\omega_{{\rm N}}}=\frac{\sqrt{8}\pi}{\omega_{{\rm CPC}}R_{{\rm int}}\sqrt{K}}e^{\frac{1}{2b}\left(\hat{L}^{b}-1\right)}\hat{L}^{1-\frac{b}{2}}\left(b^{2}-3b\hat{L}^{b}+\hat{L}^{2b}-4\right)^{-\frac{1}{2}}.
\end{equation}
\end{linenomath*}

Defining
\begin{linenomath*}
\begin{equation}
G\left(\hat{L},b\right)=\sqrt{8}\pi e^{\frac{1}{2b}\left(\hat{L}^{b}-1\right)}\hat{L}^{1-\frac{b}{2}}\left(b^{2}-3b\hat{L}^{b}+\hat{L}^{2b}-4\right)^{-\frac{1}{2}},
\end{equation}
\end{linenomath*}
we can plot how $G$ changes with $\hat{L}$ and with $b$ (Fig.~\ref{fig:freq res}).
It can be seen that for far CPCs, the oscillation period would be
extremely long, where this analysis probably does not capture the
leading order motion, while too small $L$ values might lead
to an instability that would break the periodic motion. For the Jovian
south polar CPCs, $b$ is $\sim0.78$ and $\hat{L}$ is $\sim6.5$, giving
a value of $G\approx100$ (cyan 'X' in Fig.~\ref{fig:freq res}).
The cyclones' vorticity is $\omega_{{\rm CPC}}=\sim\frac{85}{1.1\times10^{6}}\left({\rm sec}^{-1}\right)$,
while the optimization values for $K_{{\rm def}}$, $K_{{\rm trb}}$
and $R_{{\rm int}}$ are $\sim0.189,0.318,0.064$, respectively (Tab.\ \ref{tab: Optimization sets}, Set 1).
These give $T_{{\rm N}}$ of $\sim15$ months, estimating well the
12-months observed period. The remaining discrepancy might be due
to the complexities of the other cyclones, their positions in space
and their concurring oscillations, all of which are not taken into
account in this idealized derivation. 

\section{Oscillation model} \label{sec: oscillation model}

\begin{figure}
\begin{centering}
\includegraphics[width=1\columnwidth]{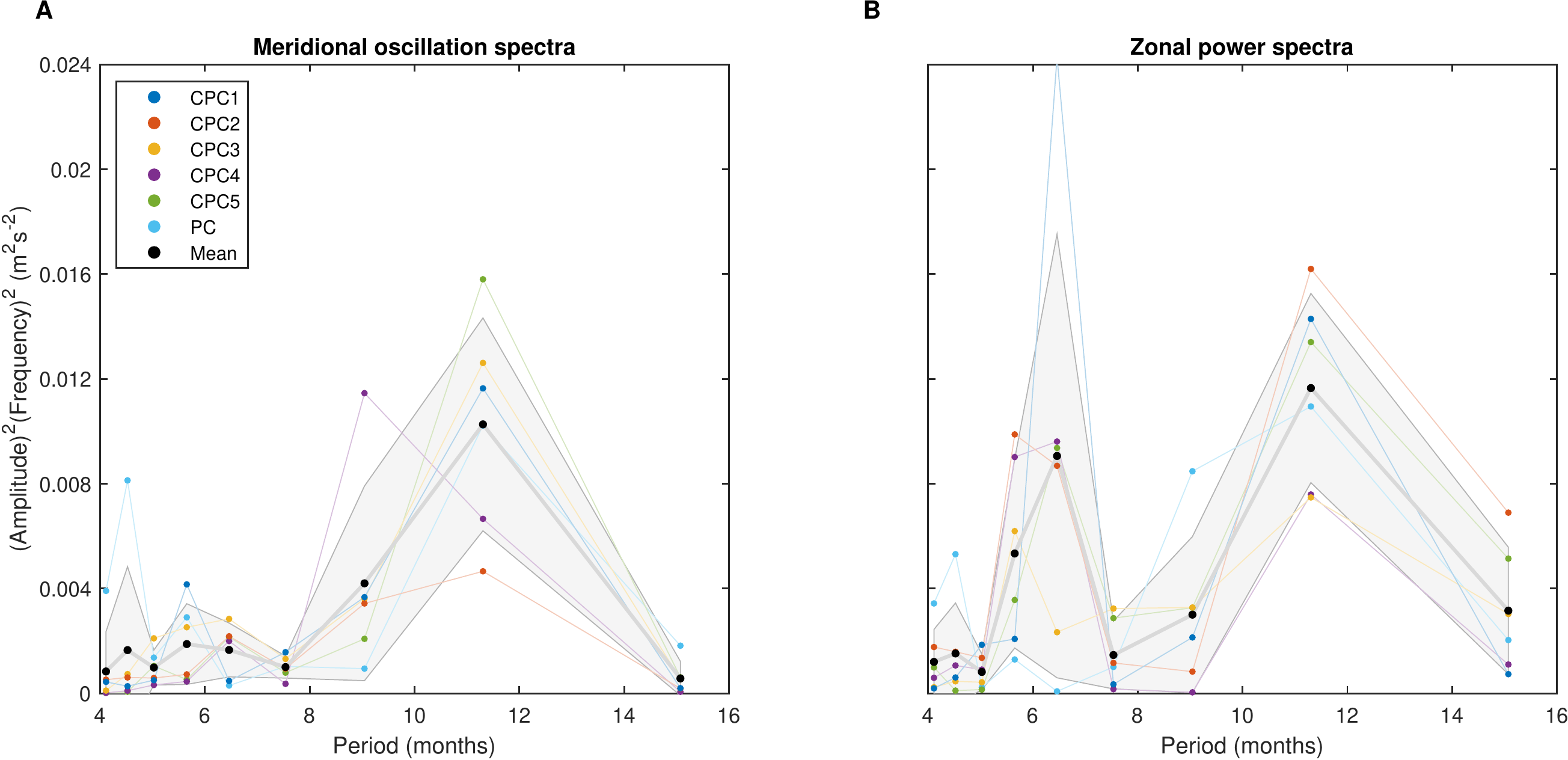}
\par\end{centering}
\caption{Energy spectra of the simulated motion. Here, panels a-b are the same as Fig.\:\ref{fig:Model results} panels b and c, but here the ordinate represents the energy of the oscillations (amplitude$^2$ frequency$^{2}$) rather than just the amplitude ($A_n$).  \label{fig: 3 power spectra}}
\end{figure}

In order to show how the underlying mechanism highlighted in this
study, i.e., the mutual rejection between cyclones due to cyclone generated
vorticity gradients and the cyclone polar attraction due to $\beta$,
produces the observed oscillation patterns, a model is created which
simulates the cyclones' motion with time according to these forces.
Here, the force calculations, described in Eqs.\ \ref{eq: Tangential velocity profile}-\ref{eq: Net force},
are used in the system of equations,
\begin{linenomath*}
\begin{equation}
\begin{array}{cccccc}
\frac{\partial u_{i}}{\partial t}= & F_{\beta i,x}+\underset{j\neq i}{\sum}F_{ji,x}, &  &  & \frac{\partial v_{i}}{\partial t}= & F_{\beta i,y}+\underset{j\neq i}{\sum}F_{ji,y},\\
\frac{\partial x_{i}}{\partial t}= & u_{i,x}, &  &  & \frac{\partial y_{i}}{\partial t}= & u_{i,y}.
\end{array}\label{eq: Model full Cartesian equations}
\end{equation}
\end{linenomath*}
For initial conditions, we use the first available location data,
which was obtained during PJ4, as 
\begin{linenomath*}
\begin{equation}
\begin{array}{cccccc}
x_{i}\left(0\right)= & x_{i,{\rm obs}}\left[{\rm PJ4}\right], &  &  & y_{i}\left(0\right)= & y_{i,{\rm obs}}\left[{\rm PJ4}\right],\end{array}
\end{equation}
\end{linenomath*}
where the subscript 'obs' refers to observational data per PJ. For
the velocity initial conditions, we simply use forward derivations.
For CPCs 1 and 2 and for the PC, we use
\begin{linenomath*}
\begin{equation}
\begin{array}{cccccc}
u_{i}\left(0\right)= & \frac{x_{i,{\rm obs}}\left[{\rm PJ5}\right]-x_{i,{\rm obs}}\left[{\rm PJ4}\right]}{53({\rm days})}, &  &  & v_{i}\left(0\right)= & \frac{y_{i,{\rm obs}}\left[{\rm PJ5}\right]-y_{i,{\rm obs}}\left[{\rm PJ4}\right]}{53({\rm days})},\end{array}
\end{equation}
\end{linenomath*}
and for CPCs 3,4 and 5, where PJ5 data is missing \citep{mura2021oscillations},
we use
\begin{linenomath*}
\begin{equation}
\begin{array}{cccccc}
u_{i}\left(0\right)= & \frac{x_{i,{\rm obs}}\left[{\rm PJ6}\right]-x_{i,{\rm obs}}\left[{\rm PJ4}\right]}{106({\rm days})}, &  &  & v_{i}\left(0\right)= & \frac{y_{i,{\rm obs}}\left[{\rm PJ6}\right]-y_{i,{\rm obs}}\left[{\rm PJ4}\right]}{106({\rm days})}.\end{array}
\end{equation}
\end{linenomath*}
From these initial conditions, the model numerically integrates Eqs.\ \ref{eq: Model full Cartesian equations}
over the duration of $45$ months ($1375$ days). The integration
time interval $dt=8({\rm hrs})$ was found to be the biggest robust
time step. Finally, For finding the unknown variables (i.e., $K_{{\rm def}},K_{{\rm trb}},R_{{\rm int}},V_{i},R_{i}$
and $b_{i}$, as defined in the text), an optimization code is performed,
which minimizes the mean differences between observed and simulated
values of $x_{i}$, $y_{i}$, and the mean differences between the
real and the simulated motion's spectral amplitudes per period. The
parameter set determined by this procedure is labeled 'Set\ 2' in
Table.\ \ref{tab: Optimization sets}, and results in the simulation
run shown in Fig.\:\ref{fig:Model results} and in movie S2.  In Fig.\:\ref{fig: 3 power spectra}, the spectra of the simulated motion is presented in terms of energy instead of metric amplitudes.

\begin{figure}
\begin{centering}
\includegraphics[width=1\columnwidth]{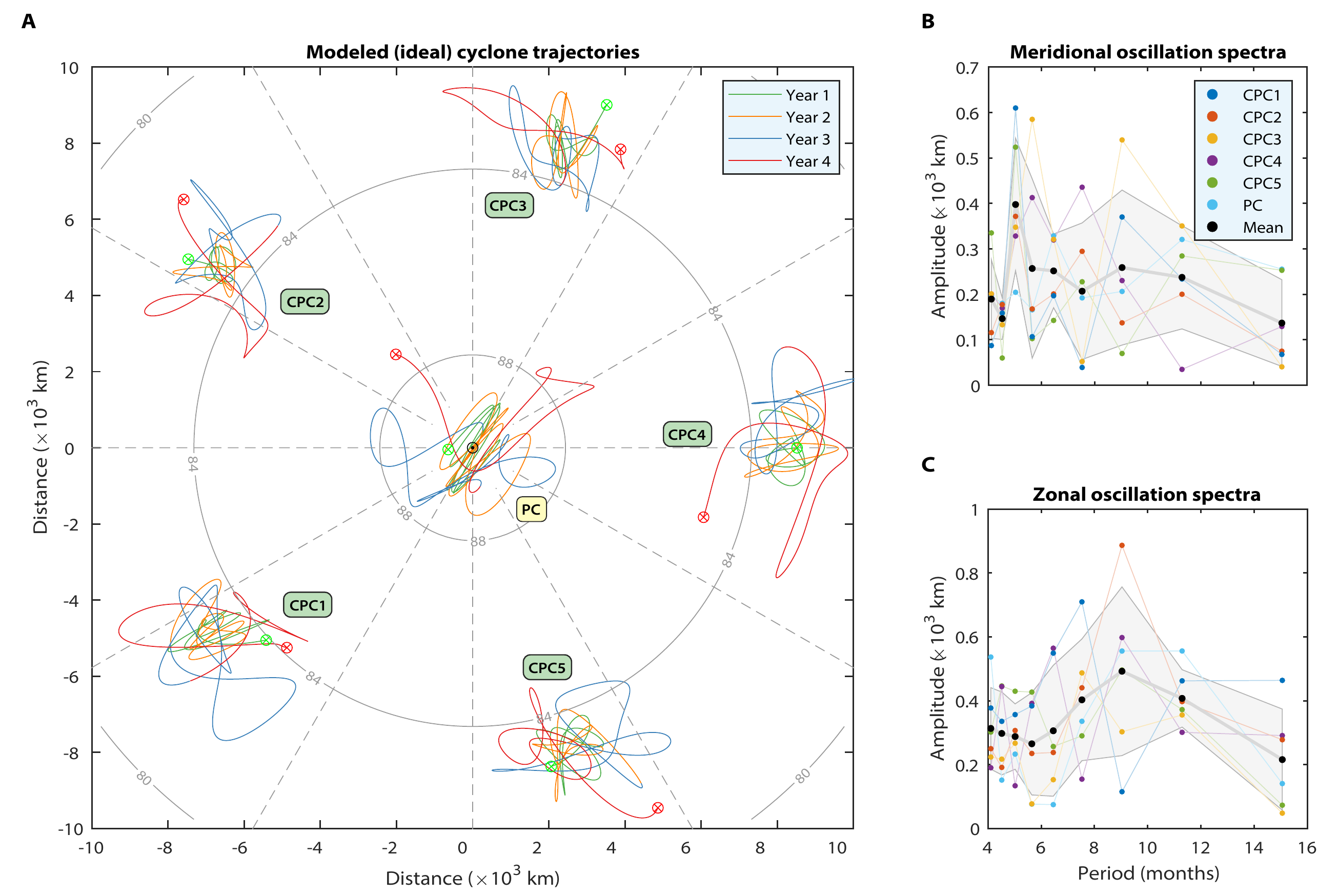}
\par\end{centering}
\caption{Ideal model simulation results. (a-c) The panels are the same as in
Fig.\:\ref{fig:Model results}. Here the cyclones are identical (Table~\ref{tab: Optimization sets},
'Ideal' case) and the initial conditions are a random perturbation
from a symmetric spread.\label{fig: Ideal model run}}
\end{figure}
\subsection{Ideal run - model robustness }

For a test of the model robustness in simulating the oscillation patterns
regardless of the optimization procedure, an 'ideal' model run is
performed. Here, all the cyclones are identical, with $V_{i}=85\ ({\rm m\ s^{-1}})$
and $R_{i}=1,100\ ({\rm km})$ as in GK21,
and with $b_{i}=0.78$. For the global coefficients, we choose $K_{{\rm def}}=0$
(no deflection), $K_{{\rm trb}}=1$ (no turbulence effect) and $R_{{\rm int}}=10\%$
(see Table.\ 1, 'Ideal' case for reference). For initial conditions,
the cyclones started at rest ($u_{i}\left(0\right),v_{i}\left(0\right)=0$),
and are spread equally along latitude $83^{\circ}$, except for the
PC, starting at $90^{\circ}$. In addition, the initial latitude of
each cyclone was perturbed by a random number in the range $\left[-1^{\circ},1^{\circ}\right]$,
and similarly in longitude by a random number in the range $\left[-8^{\circ},8^{\circ}\right]$
. These settings produced results (Fig.\ \ref{fig: Ideal model run}) similar, in essence,
to the observed patterns (Fig.\:\ref{fig:a.Data}), but with more
dispersed spectra, as might be expected from the random initial conditions.

\end{document}